\documentclass[modern]{aastex62}
\hypersetup{citecolor=blue, filecolor=blue, urlcolor=blue}
\usepackage{braket}
\usepackage{amsmath}
\usepackage{bm}
\usepackage{rotating}
\usepackage{lineno}

\newcommand{\acronym}[1]{{\small{#1}}}
\newcommand{\project}[1]{\textsl{#1}}
\newcommand{\apogee}{\project{\acronym{APOGEE}}}
\newcommand{\gaia}{\project{Gaia}}
\newcommand{\wise}{\project{\acronym{WISE}}}
\newcommand{\zmass}{\project{\acronym{2MASS}}}

\revised{\today}

\shorttitle{Stellar Abundance Maps of the Milky Way Disk}
\shortauthors{Eilers et al.}

\begin{document}\sloppy\sloppypar\raggedbottom\frenchspacing
  
\title{\textbf{Stellar Abundance Maps of the Milky Way Disk}}

\noindent
\author[0000-0003-2895-6218]{Anna-Christina~Eilers}\thanks{NASA Hubble Fellow}
\affiliation{MIT Kavli Institute for Astrophysics and Space Research, 77 Massachusetts Ave., Cambridge, MA 02139, USA}

\author[0000-0003-2866-9403]{David~W.~Hogg}
\affiliation{Center for Cosmology and Particle Physics, Department of Physics, New York University, 726 Broadway, New York, NY 10003, USA}
\affiliation{Max Planck Institute for Astronomy, K\"onigstuhl 17, 69117 Heidelberg, Germany}
\affiliation{Center for Computational Astrophysics, Flatiron Institute, 162 5th Avenue, New York, NY 10010, USA}

\author[0000-0003-4996-9069]{Hans-Walter~Rix}
\affiliation{Max Planck Institute for Astronomy, K\"onigstuhl 17, 69117 Heidelberg, Germany}

\author[0000-0001-5082-6693]{Melissa~K.~Ness}
\affiliation{Center for Computational Astrophysics, Flatiron Institute, 162 5th Avenue, New York, NY 10010, USA}
\affiliation{Department of Astronomy, Columbia University, Pupin Physics Laboratories, New York, NY 10027, USA}

\author[0000-0003-0872-7098]{Adrian~M.~Price-Whelan}
\affiliation{Center for Computational Astrophysics, Flatiron Institute, 162 5th Avenue, New York, NY 10010, USA}

\author[0000-0001-8237-5209]{Szabolcs~M\'esz\'aros}
\affiliation{ELTE E\"otv\"os Lor\'and University, Gothard Astrophysical Observatory, 9700 Szombathely, Szent Imre H. st. 112, Hungary}
\affiliation{MTA-ELTE Lend{\"u}let Milky Way Research Group, Hungary}

\author[0000-0003-4752-4365]{Christian~Nitschelm}
\affiliation{Centro de Astronom{\'i}a (CITEVA), Universidad de Antofagasta, Avenida Angamos 601, Antofagasta 1270300, Chile}

\correspondingauthor{Anna-Christina Eilers}
\email{eilers@mit.edu}

\begin{abstract}\noindent
To understand the formation of the Milky Way's prominent bar it is important to know whether stars in the bar differ in the chemical element composition of their birth material as compared to disk stars. This requires stellar abundance measurements for large samples across the Milky Way's body. Such samples, e.g.\ luminous red giant stars observed by SDSS's \apogee\ survey, will inevitably span a range of stellar parameters; as a consequence, both modelling imperfections and stellar evolution may preclude  consistent and precise estimates of their chemical composition at a level of purported bar signatures, which has left current analyses of a chemically distinct bar inconclusive. Here, we develop a new self-calibration approach to eliminate both modelling and astrophysical abundance systematics among red giant branch (RGB) stars of different luminosities (and hence surface gravity $\log g$). 
We apply our method to $48,853$ luminous \apogee\ DR16 RGB stars to construct spatial abundance maps of $20$ chemical elements near the Milky Way's mid-plane, covering Galactocentric radii of $0\,{\rm kpc}<R_{\rm GC}<20\,\rm kpc$. Our results indicate that there are no abundance variations whose geometry matches that of the bar, and that the mean abundance gradients vary smoothly and monotonically with Galactocentric radius. We confirm that the high-$\alpha$ disk is chemically homogeneous, without spatial gradients. Furthermore, we present the most precise [Fe/H] vs. $R_{\rm GC}$ gradient to date with a slope of $-0.057\pm0.001\rm~dex\,kpc^{-1}$ out to approximately $15$~kpc. 
\end{abstract}


\section{Introduction}

Stellar spectra preserve signatures of the chemical element composition of the interstellar medium from which they formed, enabling us to trace back the Galactic enrichment history \citep[e.g.][]{RixBovy2013_ARAA}. This chemical footprint provides an archaeological record of the Milky Way's composition in its past. Quantitative comparisons between observed stellar properties in orbit and chemical abundance space compared to chemo-dynamical model predictions enable us to study the Milky Way's star formation and enrichment history, since the present-day structure of our Galaxy reflects the initial conditions of the  chemical composition and dynamical state of the Milky Way during its early phases of formation \citep[e.g.][]{Fragkoudi2018, Ness2019, Frankel2018, Frankel2020}. 

Large-scale Galactic spectroscopic surveys such as \apogee\ \citep{Majewski2017, Holtzman2018, Wilson2019, apogeeDR16} as part of the \textit{Sloan Digital Sky Survey} \citep[SDSS;][]{Gunn2006, Blanton2017} enable detailed studies of the chemical composition of stellar populations over vast parts of our Galaxy. Combined with the astrometric information by ESA's \gaia\ satellite \citep{Gaia2020} providing precise phase-space information we have now the tools available to study the formation, enrichment, and evolution of our Galaxy in unprecedented detail. 




In order to derive precise chemical abundances of stars from their spectra, the \apogee\ Stellar Parameters and Chemical Abundances Pipeline \citep[ASPCAP;][]{Holtzman2015, Nidever2015, GarciaPerez2016} makes use of a multi-dimensional synthetic spectral grid that covers a wide range of fundamental stellar parameters ($\log g$, $T_{\rm eff}$, microturbulent velocity, and the abundances [M/H], [$\alpha$/M], [C/M], and [N/M]) and comprises several million individual synthetic spectra. Note that the [$\alpha$/M] abundance reported by ASPCAP denotes the average over the measured $\alpha$-elements, namely O, Mg, Si, S, Ca and Ti. 
However, decisions in the analysis pipeline may lead to model-dependent trends. Much effort has gone into calibrating and correcting for these effects by comparing for instance the derived abundances to cluster stars, which are assumed to be homogeneous \citep[e.g.][]{Bovy2016}, by applying different calibration relations to subsamples of stars, or by applying zero-point shifts to match the mean abundances to measurements in the solar neighborhood \citep[see e.g.][for details]{Holtzman2015, Holtzman2018, Nidever2015, Jonsson2018}. 
Nevertheless, small residuals and systematic trends in the derived stellar abundances from ASPCAP with respect to parameters such as the stellar surface gravity $\log g$ and effective temperature $T_{\rm eff}$, remain. These trends are a combination of expected dependencies due to stellar evolution \citep{Dotter2017}, as well as remaining systematic biases in the calibration and data processing \citep{Roederer2014, Hawkins2016, Jofre2019}.

In this work we take a sample of luminous stars on the upper end of the red giant branch (RGB) and construct a data-driven model in order to self-calibrate the stellar chemical abundances (Al, C, Ca, Ce, Co, Cr, Cu, Fe, K, Mg, Mn, N, Na, Ni, O, S, Si, Ti, P, V), aiming to correct for any trends in their abundances that depend on ``nuisance'' parameters, for which we take $\log g$. Our method removes both systematic biases and at the same time any stellar evolutionary effects with $\log g$ (we do not seek to disentangle real effects from systematic artefacts), therefore showing the photospheric abundances as if all stars were at the \textit{same evolutionary state}. 

This calibration has specific (but far-reaching) utility. While the corrected, self-calibrated measurements can not be used to investigate chemical abundance changes with stellar evolution, for Galactic archaeology, which the surveys are largely founded upon, this calibration is paramount. For example, to examine maps of abundance gradients across the galaxy -- in the context of galaxy formation -- requires the imprint of the stellar evolutionary states to be removed. Additionally, new work that uses stellar abundances as independent invariants to improve dynamical inferences about the mass distribution of the Milky Way \citep{Price-Whelan2021} also depends on an assumption that the  abundances do not change with stellar evolution.

\section{Method}\label{sec:methods}

We aim to correct any systematic effects of the stellar abundances on the ``nuisance'' parameter of surface gravity $\log g$, whether they are introduced by modelling systematics, e.g. line blending, or by stellar evolution, e.g. dredge-up. Consequently, the calibration method we apply calibrates the element abundances of all stars to the same evolutionary state.

The motivation for the structure of the model comes from the idea of \textit{causal inference} in statistics:
If measured abundances depend on both stellar evolutionary state and stellar position and velocity (stellar orbit or actions), how do we separate these dependencies?
This question would be easy to answer if all stellar evolutionary states were observed equally in all parts of the Galaxy.
However, they are not. Stars further up the red-giant branch are more luminous and hence observed over larger parts of the Galaxy disk.
Thus the \apogee\ target selection is different in different parts of the sky, which view different parts of the Galaxy.
In particular, there were different targeting considerations in the fields that cover the Milky Way bulge and bar, relative to other Galaxy-disk fields \citep[e.g.][]{Zasowski2017, apogeeDR16}.

Our model assumes that the chemical abundances of stars in the Milky Way disk depend on their birth place and epoch \citep{Sanders2015, Frankel2018}, which in turn is reflected statistically in their current orbits, quantified by the three orbital actions $J_R$, $J_z$, and $L_z$. The measured abundances may also depend on the nuisance parameter $\log g$.
Thus, we presume that the stellar abundances can be modeled as a function of orbital actions and $\log g$.
If we adopt a linear model, we can then separate the physically \textit{expected} dependencies on the orbital actions from the \textit{unexpected} dependencies of the stellar parameters on $\log g$, which we can simply subtract off. Thus we aim to fit for all dependencies of the stellar abundances, and afterwards subtract the dependencies on nuisance parameters. 

The structure of the model---fit for dependencies on both $\log g$ and on orbital actions---when we only care about removing or calibrating out the $\log g$ dependencies, flows from ideas in causal inference:
If we were to just fit for dependence of element abundances on $\log g$, we would find strong dependencies for all elements, since element abundances depend on Galactocentric position, and hence on their distance from us, thereby on luminosity (in a flux-limited sample) and thus on $\log g$.  
This effect is further compounded by variations in the \apogee{} selection function as a function of Galactic latitude and longitude, which further exacerbates the spatial variations in the $\log g$ distribution. If we simultaneously fit for dependence of abundances on both $\log g$ and orbital actions (as a physically better motivated proxy than mere Galactic location), the regression can separate the effects coming from $\log g$ and from Galactic orbit.
Then the $\log g$ dependencies we find have effectively accounted properly for the confounding effect of Galactic orbit.
When we calibrate or adjust the abundances, we adjust them using only the $\log g$ part of the model.

To this end, for each star in our data set (see \S~\ref{sec:data}) we construct a feature vector $\vec{y}$ containing linear and quadratic terms of the orbital actions $J_R$, $J_z$, and $L_z$, which are the parameters on which the ``true'' stellar abundances are expected to depend, and the nuisance parameter $\log g$. 
Note that in practice we subtract an arbitrarily chosen offset of 2.2 from all $\log g$ values, i.e. we take $\log\tilde g=\log g-2.2$. This implies that our model calibrates all stellar abundances to the same evolutionary state at $\log g\approx2.2$, i.e. just above the red clump. 

We chose the $D$-dimensional feature vector $\Vec{y}$ to be a second order polynomial, i.e. 
\begin{align}
    \vec{y} = [1, J_R, J_z, L_z, J_R^2, J_RJ_z, J_RL_z, J_z^2, J_zL_z, L_z^2, \log \tilde g, \log^2 \tilde g],  \label{eq:feature}
\end{align}
where $D=12$ in this case. 
The matrix $\bf Y$ contains the feature vectors for each of the $N$ stars in our sample, i.e. it has a shape $D\times N$. For each element abundance ${\rm [X/Fe]}_{\rm DR16}$ reported in \apogee\ DR16 we now find a $D$-dimensional coefficient vector $\Vec{\vartheta}$ containing the coefficients for each feature, which minimizes the Euclidean 2-norm, i.e.
\begin{equation}
   \hat{\vec{\vartheta}}\longleftarrow\underset{\vec{\vartheta}}{\rm argmin}\left[\frac{\left({\rm [X/Fe]}_{\rm DR16} - {\bf Y}\cdot\Vec{\vartheta}\right)^2}{\sigma^2_{\rm [X/Fe]}}\right], \label{eq:optimization}
\end{equation}
where $\sigma_{\rm [X/Fe]}$ denotes the reported uncertainty on the abundance measurements. 

The subset of $\Vec{y}$ that contains the nuisance parameter $\log \tilde{g}$ that we aim to eliminate is: 
\begin{align}
    \Vec{y}_{\rm nuis} = [\log\tilde g, \log^2 \tilde g] \label{eq:nuis}
\end{align}
In order to obtain the self-calibrated stellar abundances ${\rm [X/Fe]}_{\rm new}$ we simply subtract any dependencies on these nuisance parameters, i.e. 
\begin{align}
   {\rm [X/Fe]}_{\rm new} = {\rm [X/Fe]}_{\rm DR16} - {\bf Y}_{\rm nuis} \cdot \hat{\Vec{\vartheta}}_{\rm nuis}. 
\end{align}
Note, that this method does not deliver an \textit{absolute} abundance calibration, but only a relative calibration providing us with the stellar abundance \textit{gradients} across the Milky Way disk. 

It is also worth noting that the only nuisance parameters that our model explicitly accounts for are $\log\tilde g$ and $\log^2\tilde g$ (Eqn.~\ref{eq:nuis}). Hence the corrected, self-calibrated abundances depend on their stellar orbits, but no longer on their surface gravity \textit{at any given orbit}. The abundances can also vary with respect to other quantities that correlate with their orbit, such as for instance the stellar age, i.e. $\tau_{\rm age}=f(J_R, J_z, L_z)$ \citep[e.g.][]{Beane2018, Frankel2018, Frankel2020}. Therefore, our model controls not just for actions but for any stellar properties that physically correlate with actions. 

\section{Data}\label{sec:data}

Our analysis is based on stars on the upper red giant branch (RGB), which are ideally suited to map the Milky Way disk over large Galactocentric radii due to their high luminosities. However, due to the limited precision of \gaia's astrometric parallaxes beyond a few kpc distance from the Sun, we first need an alternate way to estimate precise parallaxes for these RGB stars, in order to calculate the orbital actions required in Eqn.~\ref{eq:feature}. 

To this end, we previously developed a data-driven model assuming that the RGB stars are dust-correctable, standardizable candles \citep{HER2019}. This implies that we can infer their distance modulus -- and thus their parallax -- from their spectroscopic and photometric features. This model is in detail described in \citet{HER2019} and will be summarized in \S~\ref{sec:spec_phot} briefly. In \S~\ref{sec:actions} we will calculate the orbital actions for the RGB stars within the Milky Way disk making use of the spectro-photometric parallax estimates. 

\subsection{Spectro-photometric parallax estimates}\label{sec:spec_phot}

Following \citet{HER2019} we chose a model for the stellar parallax that employs a purely linear function of spectral pixel intensities from \apogee\ DR16 \citep{Majewski2017, apogeeDR16}, as well as multi-band photometry from \gaia\ eDR3 \citep{Gaia2020}, \zmass\ \citep{2MASS}, and \wise\ \citep{WISE}. Due to the limited flexibility of this linear model we restrict ourselves to a confined region of the stellar parameter space and analyze only stars with very low surface-gravity, i.e. $\log g \leq 2.2$, which selects stars that are more luminous than the red clump. We will now briefly recap the main points of this data-driven model and note where updates to the original model presented in \citet{HER2019} have been made. For more details, however, we refer the reader to the original paper. 

We require complete photometric and spectroscopic information for all RGB stars, which results in a parent sample containing $79,529$ stars with $\log g \leq 2.2$. The model for the parallax is expressed by
\begin{equation}
    \varpi^{\rm(a)}_n = \exp\left(\vec{\theta}\cdot \vec{x}_n\right) + \rm noise, \label{eq:parallax}
\end{equation}
where $\varpi^{\rm(a)}_n$ is \gaia's astrometric parallax measurement of star $n$. In practice we add a constant offset to all astrometric parallax measurements of $\Delta \varpi^{\rm(a)} = 17~\rm \mu as$, corresponding to the reported median parallax bias in eDR3 \citep{Lindegren2021}. Our model thus postulates that the logarithm of the true parallax can be expressed as a linear combination of the components of a $D$-dimensional feature vector $\vec{x}_n$ and the $D$-dimensional coefficient vector $\vec{\theta}$. The feature vector $\vec{x}_n$ here contains the spectroscopic as well as photometric features for each star $n$, and thus consists of approximately $D\approx 7400$ features. Note that since parallaxes cannot be negative by definition, we model the logarithm of the parallax. 

The log-likelihood function can then be expressed as
\begin{equation}
    \ln \mathcal{L} = -\frac{1}{2}\chi^2(\vec{\theta})=-\sum_{n=1}^N\frac{\left[\varpi^{\rm(a)}_n - \exp(\vec{\theta}\cdot \vec{x}_n)\right]^2}{2\sigma^{\rm(a)2}_n},   
\end{equation}
where $\sigma^{\rm(a)}_n$ denotes the uncertainty on \gaia's astrometric parallax measurement. 

Furthermore, we assume that many entries in the coefficient vector $\vec{\theta}$ will be zero, which is known as the sparsity assumption. Thus we apply a regularization and optimize the regularized objective function, i.e. 
\begin{equation}
    \hat{\vec{\theta}} \longleftarrow\underset{\vec{\theta}}{\rm argmin}\left[\frac{1}{2}\chi^2(\vec{\theta})+\lambda\|P\cdot\vec{\theta}\|_1^1\right],  
\end{equation}
where $\lambda$ is the regularization parameter, and $P$ is a projection operator that selects only those components that belong to the \apogee\ spectral pixels. 

For the optimization we split our parent sample randomly in two distinct sub-samples $A$ and $B$, which are used as a training and validation set (and vice versa), respectively. The value of $\lambda$ is set to $\lambda=160$ via cross-validation of the $A/B$-split sample. 

Making use of the optimized coefficient vector $\hat{\vec{\theta}}$ we can now infer the spectro-photometric parallax estimates $\varpi^{\rm (sp)}_m$ for each star $m$ in the validation set via 
\begin{equation}
    \varpi^{\rm (sp)}_m \longleftarrow\exp(\hat{\vec{\theta}}\cdot \vec{x}_m). 
\end{equation}

We have made two improvements compared to the model presented previously in \citet{HER2019}, namely 
\begin{enumerate}
    \item The \gaia\ parallaxes and photometry have been updated with the new data release eDR3. 
    \item We make use of the updated and enlarged \apogee\ DR16 catalog \citep{apogeeDR16}, which includes spectra not only from the Northern hemisphere observed with the Apache Point Observatory in New Mexico (APO), but also from the South taken at the Las Campanas Observatory in Chile (LCO). While both telescopes have similar technical properties (e.g. mirror size, spectral resolution, wavelength coverage), there remain differences in the density of the applied fibers (i.e. 300 fibers per $7~\rm deg^2$ plate for APO or $3.5~\rm deg^2$ at LCO), which could cause mild discrepancies in the extracted spectra. Hence, we add the first five moments of the line-spread function (LSF) to our model, which in practice simply adds five additional terms to the feature vector $\vec{x}_n$ in Eqn.~\ref{eq:parallax}. 
\end{enumerate}
This data-driven model provides spectro-photometric parallax estimates with approximately $9\%$ uncertainties \citep{HER2019}, which results in distance estimates that are more than twice as precise as \gaia's predictions at heliocentric distances of $\gtrsim 3$~kpc ($\gtrsim 1$~kpc) for stars with $G\sim 12$~mag ($G\sim14$~mag). These precise distance estimates enable us to create global maps of the Milky Way out to large Galactocentric distances of up to $25$~kpc \citep{HER2019, Eilers2020}. 

\begin{figure}
    \centering
    \includegraphics[width=\textwidth]{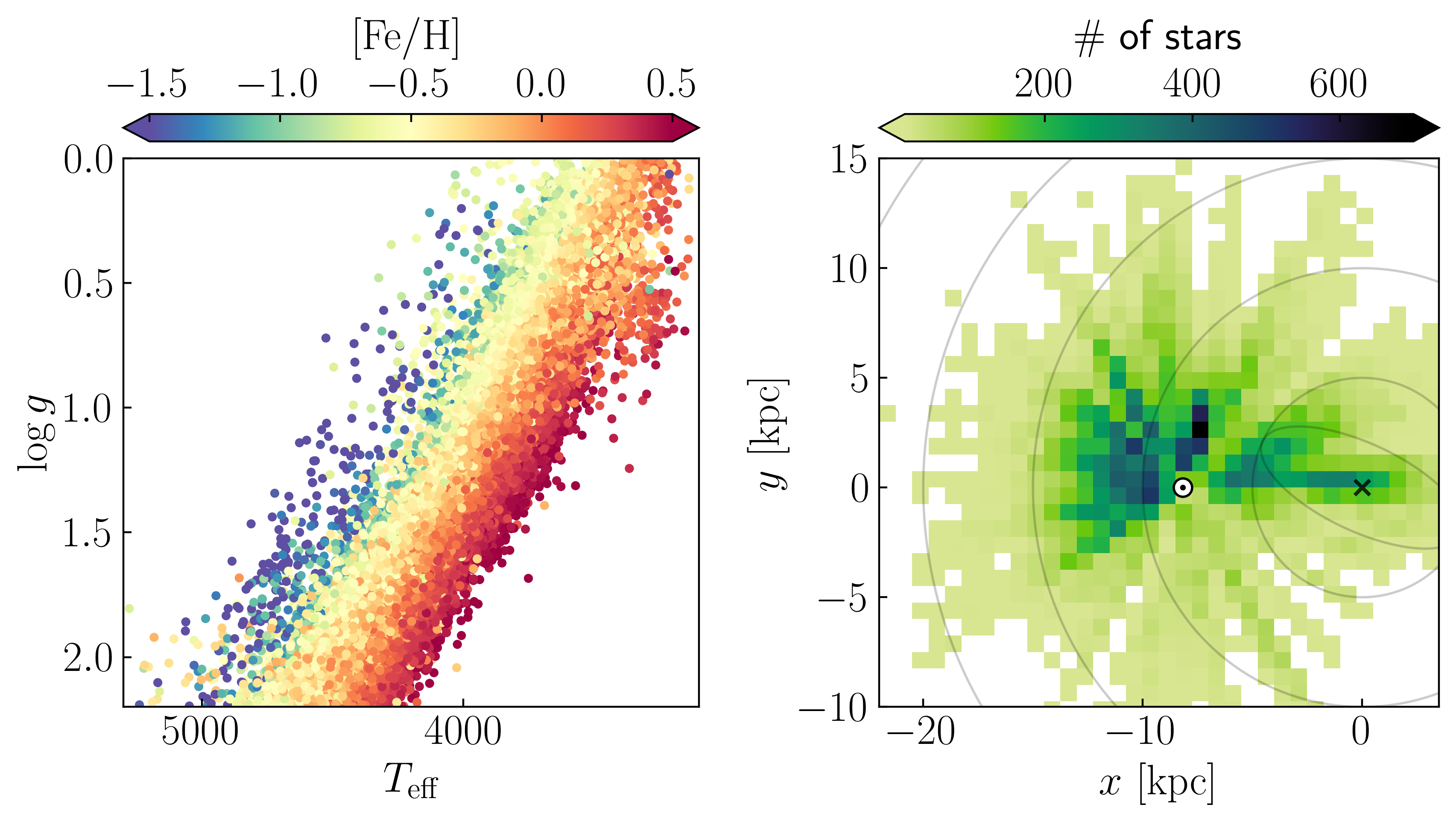}
    \caption{Sample of $48,853$ RGB stars ($\log g\leq2.2$) within the Milky Way disk used for this analysis. \textit{Left:} Distribution of surface gravity $\log g$ v.\ effective temperature $T_{\rm eff}$, colored by the metallicity [Fe/H] from \apogee\ DR16. \textit{Right:} Galactic map colored by the number of stars in each bin of $\Delta x = \Delta y= 0.75~\rm kpc$. The Galactic center and position of the Sun are indicated at their location of $(x,y) = (0,0)$, and $(x,y) = (-8.178~{\rm kpc},0)$, respectively. }
    \label{fig:sample}
\end{figure}

In this paper we present an analysis of the stellar abundances within the Milky Way disk which makes use of a subset of this data set, i.e. $48,853$ RGB stars confined to the Galactic disk. Thus we chose stars with a height above the Galactic plane of $|z_{\rm GC}|\leq 1$~kpc or within a $6^{\circ}$ wedge starting from the Galactic center to account for disk flaring. Furthermore, we require $v_z<100\,\rm km\,s^{-1}$ in order to exclude halo stars. Additionally, we exclude all stars that belong to the globular cluster $\omega$Cen, which removes another $77$ stars from our sample. Our data sample of Milky Way disk stars is presented in Fig.~\ref{fig:sample}. 

By inverting the spectro-photometric parallax estimates to infer the stellar distances we can now create large scale maps of the Milky Way disk. We transform all stars to the Galactocentric cylindrical coordinate frame making use of the barycentric radial velocities from \apogee\ DR16 and proper motions delivered from \gaia\ eDR3. The Galactocentric azimuth angle $\varphi$ is measured from the centre--anticentre line with $\varphi$ increasing counter--clockwise, i.e. in the opposite direction of Galactic rotation. We assume a distance from the Sun to the Galactic center of $R_{\odot}\approx8.178$~kpc \citep{gravity2019}, a height of the Sun above the Galactic plane of $z_{\odot}\approx 0.025$~kpc \citep{Juric2008}, and the Galactocentric velocity components of the Sun $v_{\odot,\,x}\approx-11.1\,\rm km\,s^{-1}$, $v_{\odot,\,y}\approx245.8\,\rm km\,s^{-1}$, and $v_{\odot,\,z}\approx7.8\,\rm km\,s^{-1}$, which have been derived from the proper motions of $\rm Sgr~A^{*}$ \citep{ReidBrunthaler2004}.

\subsection{Orbital Actions}\label{sec:actions}

The chemical composition of stars is expected to depend on a star's orbit, or consequently position within the Galactic disk, which is the fundamental assumption of our self-calibration model described in \S~\ref{sec:methods}. The orbital actions ($J_R$, $L_z$, $J_z$) are a powerful tool to characterize the orbits of stars in the Galactic gravitational potential. In an axisymmetric gravitational potential the actions are independent integrals of motion and quantify the amount of oscillation of a star around its orbit \citep[e.g.][]{BinneyTremaine, Trick2019}. 
Assuming an axisymmetric gravitational potential derived in \citet{Eilers2019} we can calculate the orbital actions for all stars in our data set, i.e. the angular momentum $L_z$, the radial action $J_R$, as well as the vertical action $J_z$, using the \texttt{gala} package \citep{gala}. 

We restrict the training set, which is used to optimize the coefficient vector $\hat{\vec{\vartheta}}$ in Eqn.~\ref{eq:optimization} to those stars that have reliably estimated actions. To this end we remove all stars from the training set of the model, for which the calculation of the actions failed, or that represent strong outliers in the action distribution. Hence our training set consists of a subset of the Milky Way disk stars, i.e. $43,196$ RGB stars with $0~{\rm kpc\,km\,s^{-1}}< J_R < 400~{\rm kpc\,km\,s^{-1}}$, $-5000~{\rm kpc\,km\,s^{-1}}<L_z< 1000~{\rm kpc\,km\,s^{-1}}$, and $0~{\rm kpc\,km\,s^{-1}}<J_z< 100~{\rm kpc\,km\,s^{-1}}$. We furthermore remove stars at $x>0$, i.e. stars behind the Galactic bulge, from the training set, since their distance estimates might be biased due to crowding. 

Note that in the presence of a strong Galactic bar the orbital actions for stars in the inner part of our Galaxy are no longer well-defined, since actions are only defined in an axisymmetric gravitational potential. We have thus confirmed that the applied self-calibration model delivers nearly identical results if stars at Galactocentric radii  $R<5$~kpc are completely excluded from the training process of the model, i.e. the optimization of the coefficient vector $\hat{\vec{\vartheta}}$ in Eqn.~\ref{eq:optimization}. \\

\section{Results}

In this section we show that our model eliminates (spurious) dependencies on $\log g$ and present the self-calibrated stellar abundance gradients (\S~\ref{sec:gradients}), as well as abundance maps of the Galactic disk (\S~\ref{sec:maps}). 

\begin{figure}
    \centering
    \includegraphics[width=\textwidth]{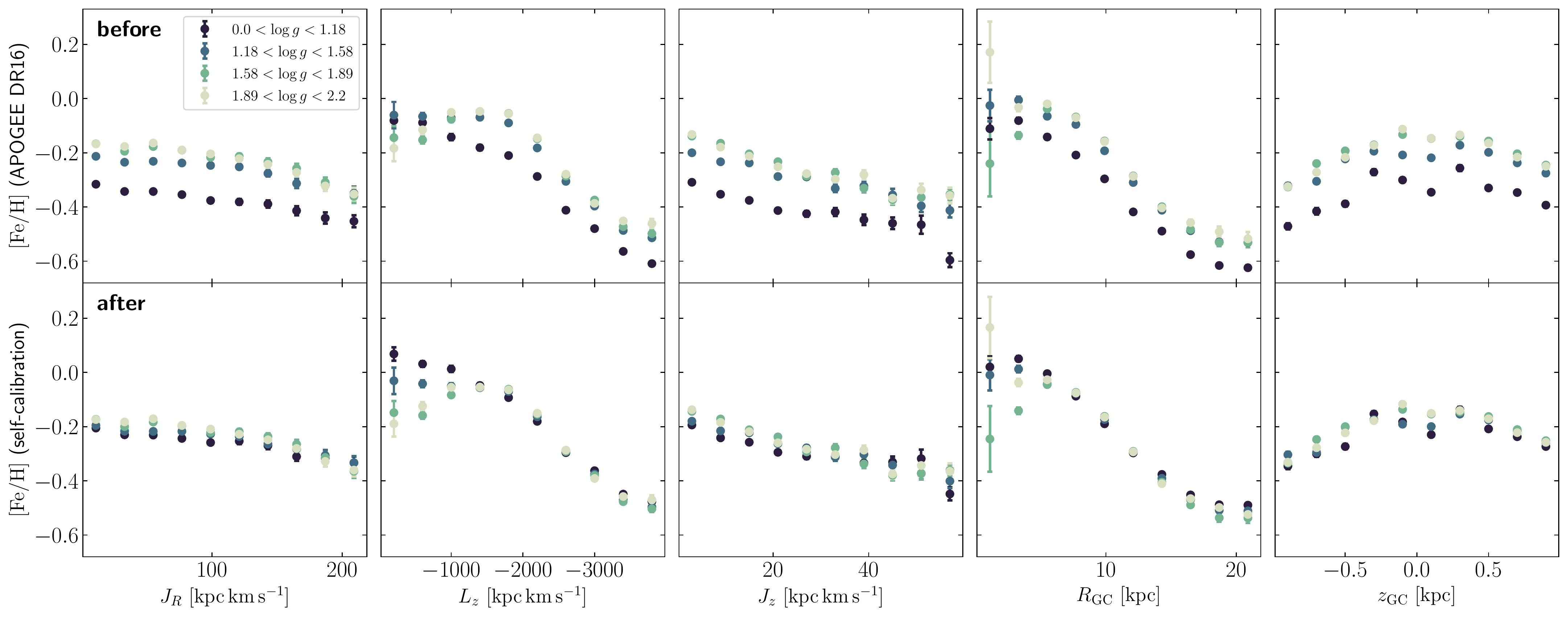}
    \caption{Stellar abundance gradients for [Fe/H] before (\textit{upper panels}) and after (\textit{lower panels}) applying the self-calibration. We bin all stars in the sample into four bins of $\log g$, each containing an equal number of stars. The stellar abundances are shown as a function of the three actions $J_R$, $J_z$, and $L_z$ that are also fit parameters (\textit{three leftmost panels}), as well as the Galactocentric radius $R_{\rm GC}$ and height $z_{\rm GC}$ above the plane (\textit{fourth and fifth panels}). While we see clear systematic trends with $\log g$ in the \apogee\ DR16 abundances, these dependencies are mostly resolved after the self-calibration, although some small effects remain, in particular at small Galactic radii. } 
    \label{fig:feh}
\end{figure}

\begin{figure}
    \centering
    \includegraphics[width=\textwidth]{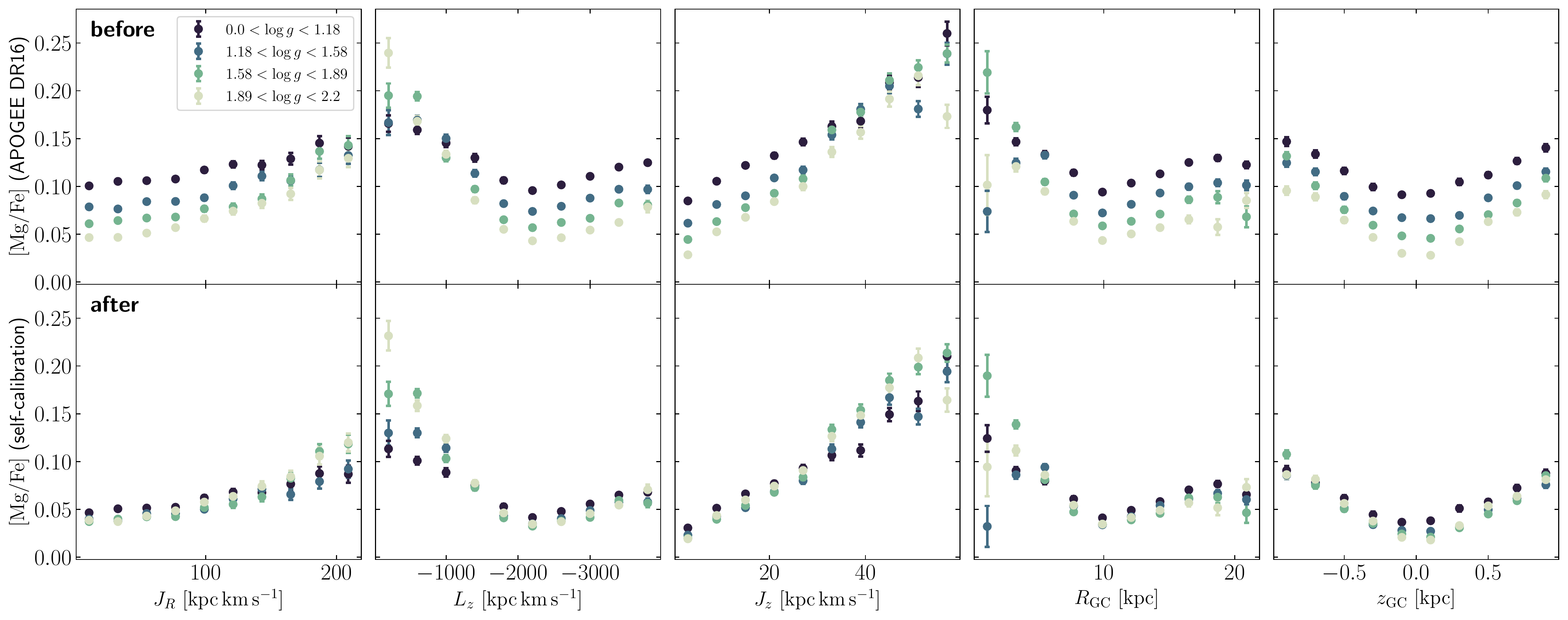}
    \caption{Same as Fig.~\ref{fig:feh} for [Mg/Fe]. }
    \label{fig:mgfe}
\end{figure}

\begin{figure}
    \centering
    \includegraphics[width=\textwidth]{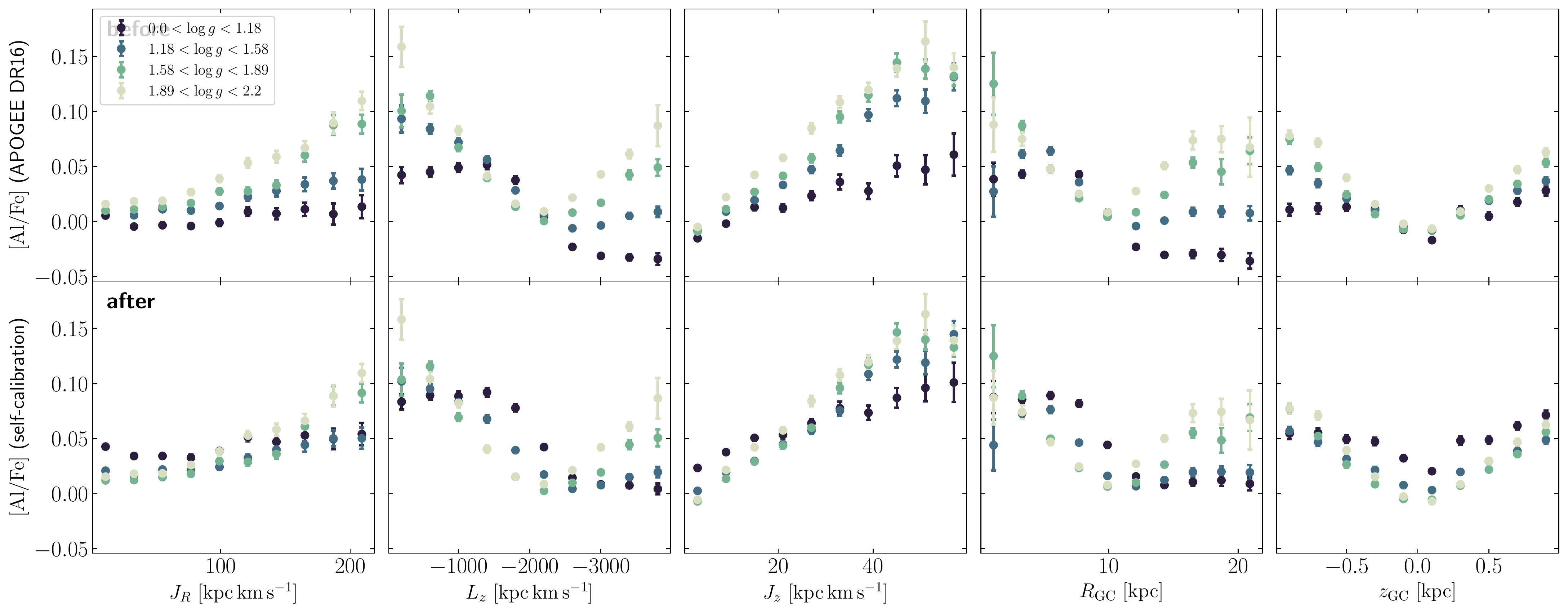}
    \caption{Same as Fig.~\ref{fig:feh} for [Al/Fe]. For this element, obvious systematic trends with $\log g$ remain visible even after applying the self-calibration model. }
    \label{fig:alfe}
\end{figure}

\subsection{Stellar abundance gradients}\label{sec:gradients}

The stellar abundances depend on the integrals of motions of their orbit, i.e. $J_R$, $J_z$, and $L_z$, and thus also on the Galactocentric radius $R_{\rm GC}$ and height above the Galactic plane $z_{\rm GC}$. In Fig.~\ref{fig:feh} we show the abundance gradients for the metallicity [Fe/H] for the labels from the \apogee\ DR16 catalog before calibration (top panels), as well as for the newly determined abundances after applying the self-calibration (bottom panels). 

It is evident that the systematic dependencies on the nuisance parameter $\log g$ that are easily visible in the \apogee\ DR16 abundances significantly decrease or are eliminated completely after applying the self-calibration model. Note that while our model uses the orbital actions to calibrate the abundances (Eqn.~\ref{eq:feature}), the dependencies on $R_{\rm GC}$ and $z_{\rm GC}$ are not explicitly part of the calibration procedure and enter only implicitly as the positions of the stars are used in the calculation of the orbital actions. Nevertheless the systematic trends of the abundances with $\log g$ as a function of $R_{\rm GC}$ and $z_{\rm GC}$ decrease (see panels in the fourth and fifth column of Fig.~\ref{fig:feh}). This also becomes evident in Fig.~\ref{fig:mgfe}, which is analogous to Fig.~\ref{fig:feh} but for the magnesium abundance [Mg/Fe]. 

However, some systematic dependencies on $\log g$ nevertheless persist even after the calibration, as shown in Fig.~\ref{fig:alfe} for the aluminium abundance [Al/Fe], but also visible in Fig.~\ref{fig:feh} and \ref{fig:mgfe} at very small Galactocentric radii for instance. This is likely due to the limited flexibility of the applied quadratic model (see Eqn.~\ref{eq:feature}) used to perform the self-calibration, as well as due to possible further systematic dependencies on nuisance parameters other than $\log g$. 

Our final catalog of self-calibrated abundance estimates as well as updated spectro-photometric parallaxes is publicly  available\footnote{The catalog can be downloaded here: \href{https://doi.org/10.5281/zenodo.5963324}{https://doi.org/10.5281/zenodo.5963324}. It contains all entries from the Apogee DR16 catalog \citep{apogeeDR16}, as well as new entries for the self-calibrated abundances (columns are called \texttt{X\_FE\_NEW} and refer to element [X/Fe]) and spectro-photometric parallax estimates (the column name is \texttt{spec\_parallax} with the corresponding uncertainty \texttt{spec\_parallax\_err}). }.

\subsection{Galactic maps of stellar abundances}\label{sec:maps}

\begin{figure}[!t]
    \centering
    \includegraphics[width=\textwidth]{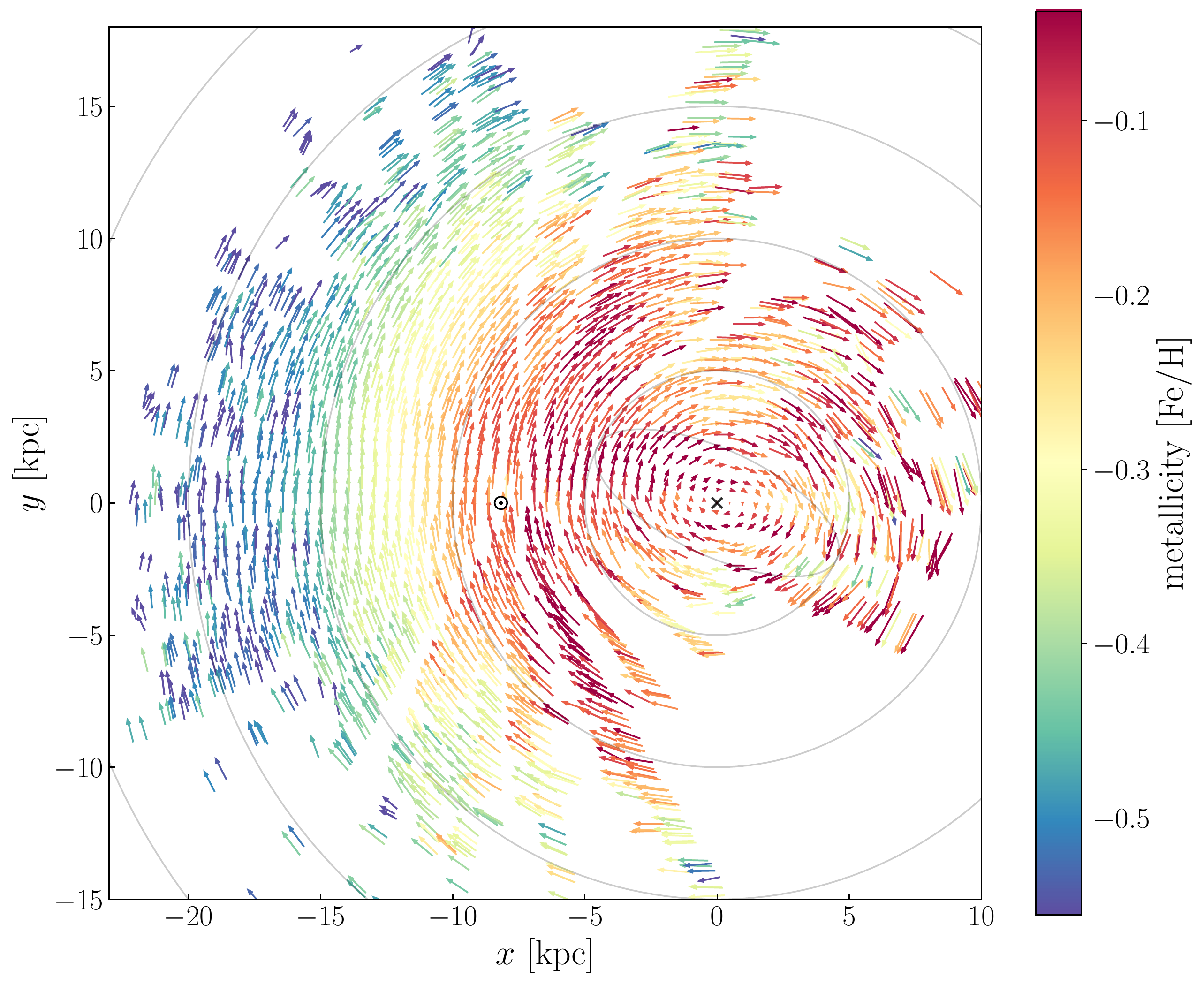}
    \caption{Face-on map of the Galactic disk within $|z|<1$~kpc showing the rotation of stars around the Galactic center located at $(x, y) = (0, 0)$. Stars are binned in spatial bins of $\Delta x=\Delta y = 0.5~\rm kpc$ and colored by the mean self-calibrated metallicity [Fe/H] of stars within that bin, while arrows point in the mean direction of motion within the plane. The Sun's position at $(x,y)=(-8.178\,{\rm kpc}, 0)$ is indicated by $\odot$, and the Galactic center is located at $(x,y)=(0, 0)$. Thin grey lines show Galactocentric rings in steps of $\Delta R=5$~kpc, as well as the outline of the Galactic bar with a half-length of the major axis of $5$~kpc and an axis ratio of $0.4$, angled at $25^\circ$ from our line of sigh to the Galactic Center.}
    \label{fig:dist1}
\end{figure}

\begin{figure}
    \centering
    \includegraphics[width=\textwidth]{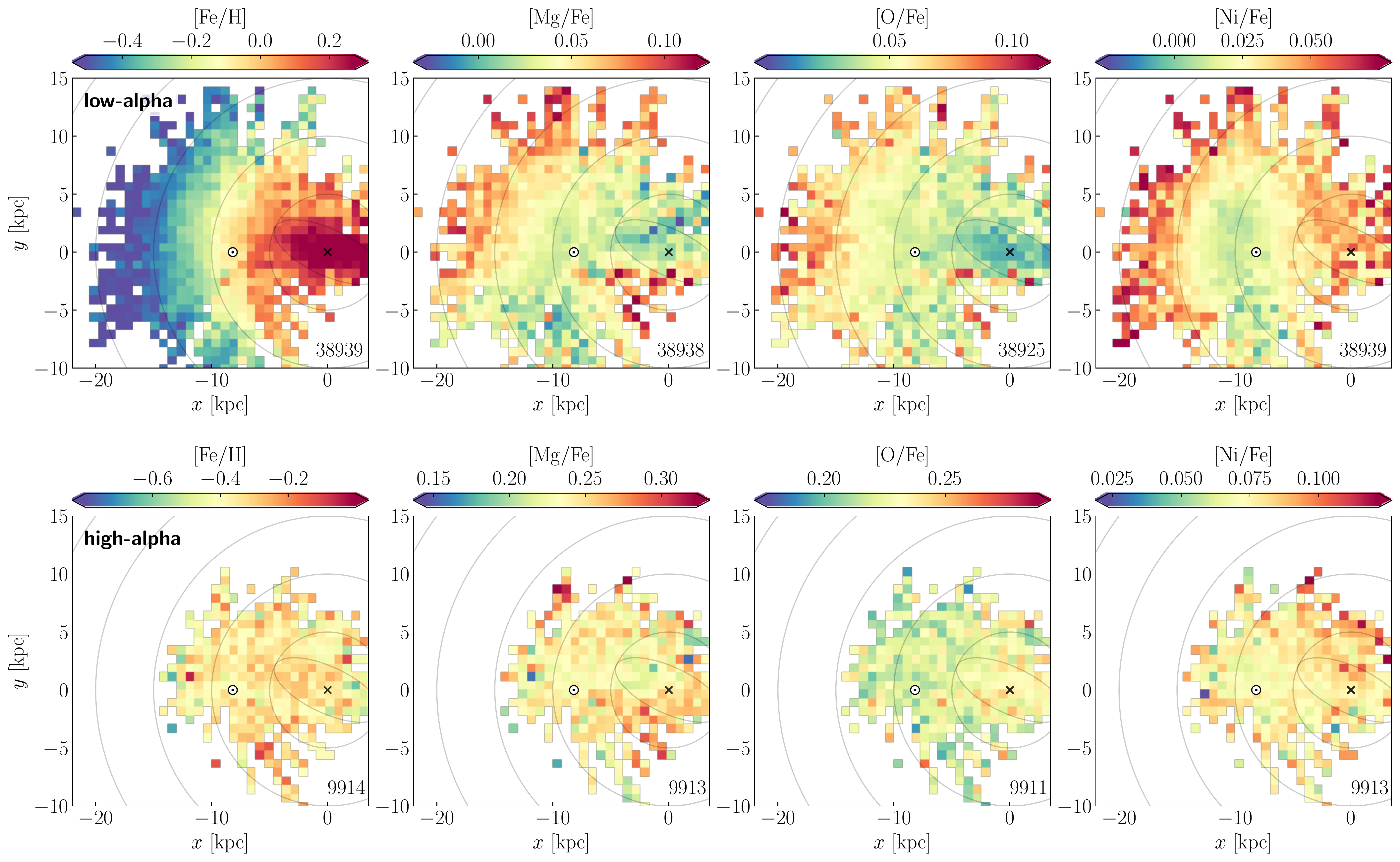}
    \caption{Maps of the Galactic disk (within $|z|<1~\rm kpc$ or a $6^\circ$ wedge) colored by four different chemical abundances after applying the self-calibration for low-$\alpha$ stars, i.e. $[\alpha/{\rm M}] < 0.12$ \textit{(top panels)}, as well as high-$\alpha$ stars, i.e. $[\alpha/{\rm M}] > 0.12$ \textit{(bottom panels)}. We show the median abundances of all stars in spatial bins of $\Delta x = \Delta y = 0.75$~kpc, whenever a bin contains more than $3$ stars. The colorbar ranges between the $7$th to $93$rd percentile of median abundances in each panel. The Galactic center is located at $(x,y)=(0,0)$. The position of the Sun is marked by $\odot$. The number of stars in each map is noted in the lower right corner of each plot. Maps for all other stellar abundances are shown in Fig.~\ref{fig:maps_lo} and \ref{fig:maps_hi} in Appendix~\ref{app:all_maps}. 
    \label{fig:maps}}
\end{figure}

\begin{figure}[!]
    \centering
    \includegraphics[width=\textwidth]{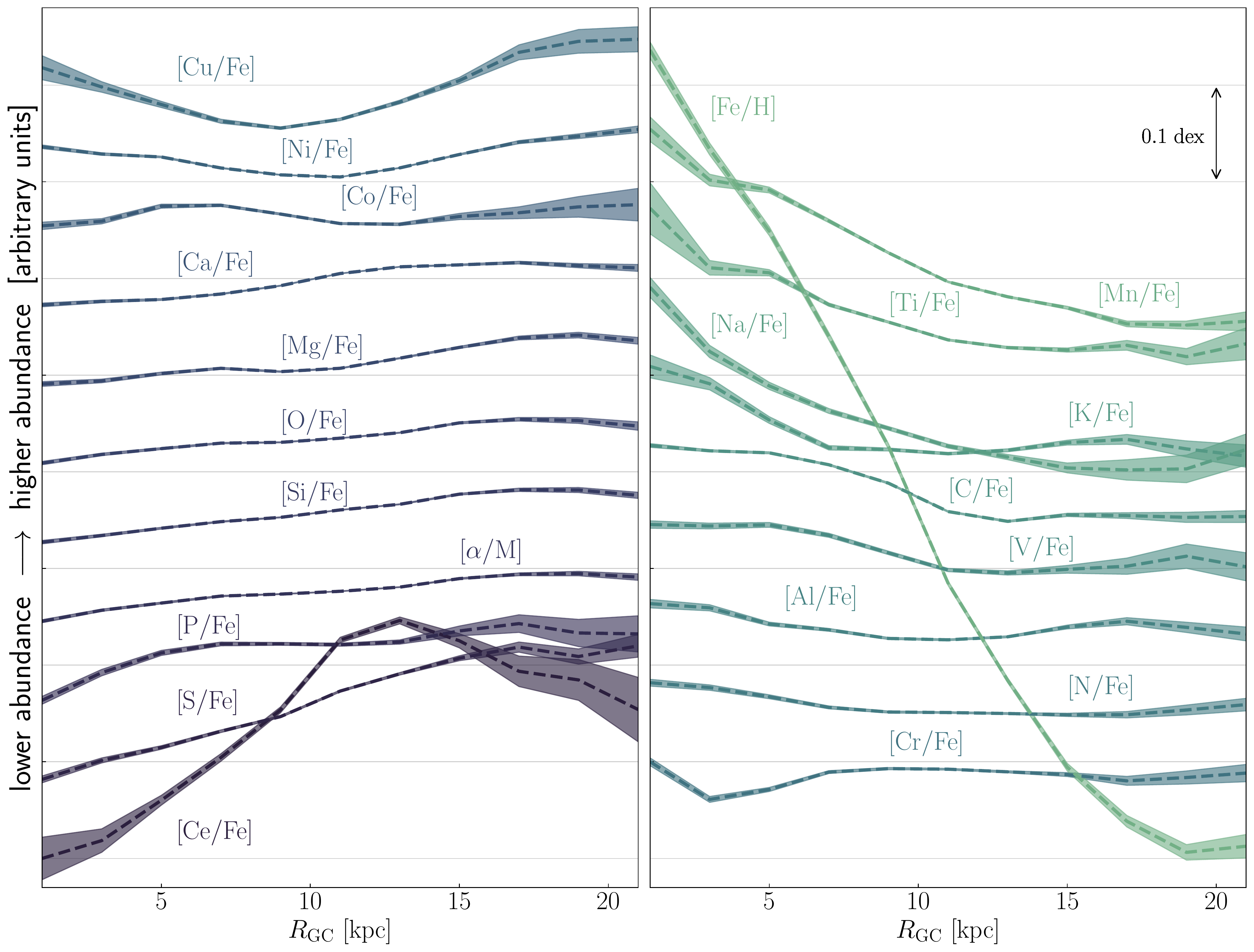}
    \caption{Median abundance gradients for the low-alpha (i.e. $[\alpha/{\rm M}]<0.12$) population as a function of Galactocentric radius $R_{\rm GC}$ with uncertainties on the median, i.e. $1.253\,\sigma/\sqrt{N}$. The abundances are sorted by the slope of their gradients -- ranging from overall positive slopes (purple, \textit{left panel}) to predominantly negative slopes (green, \textit{right panel}). An arbitrary offset is added to all gradients for better visibility. The magnitude of the gradients can be read off by the grey lines which are spaced by $\Delta=0.1\rm\, dex$. All abundances gradients can also be found in Table~\ref{tab:gradients}. 
    \label{fig:gradients}}
\end{figure}

After removing systematic and evolutionary effects on the stellar abundances by means of the self-calibration we can now map the Galactic disk colored by different element abundances to study their evolution across the disk. As an example we show in Fig.~\ref{fig:dist1} a face-on map of the Galactic disk colored by the mean self-calibrated stellar metallicity [Fe/H] in spatial bins of $\Delta x=\Delta y=0.5~\rm kpc$. The arrows indicate the mean motion of the stars around the Galactic center. 

In Fig.~\ref{fig:maps} we show maps of the Galactic disk for four abundances ([Fe/H], [Mg/Fe], [O/Fe], [Ni/Fe]) 
split between low- and high-$\alpha$ stars. We use the self-calibrated metallicity [$\alpha$/M] to split the sample, i.e. [$\alpha$/M]~$<0.12$ and [$\alpha$/M]~$>0.12$, respectively. Maps of all remaining abundances measured by \apogee\ DR16 for which we applied our self-calibration approach are shown in Fig.~\ref{fig:maps_lo} and \ref{fig:maps_hi} in Appendix~\ref{app:all_maps}, while the differences between the self-calibrated and the reported \apogee\ DR16 abundances are shown in Fig.~\ref{fig:maps_lo_diff} and \ref{fig:maps_hi_diff} in Appendix~\ref{app:apogee}. We find that all abundances evolve smoothly with Galactocentric radius, as also shown in Fig.~\ref{fig:gradients}. The abundance gradients are tabulated in Table~\ref{tab:gradients} to \ref{tab:gradients3} in Appendix~\ref{app:all_maps}. We do not see any evidence for chemically distinct features, nor do stars in the Galactic bulge and bar region appear to have a distinct abundance pattern. 

\begin{figure}[!]
    \centering
    \includegraphics[width=\textwidth]{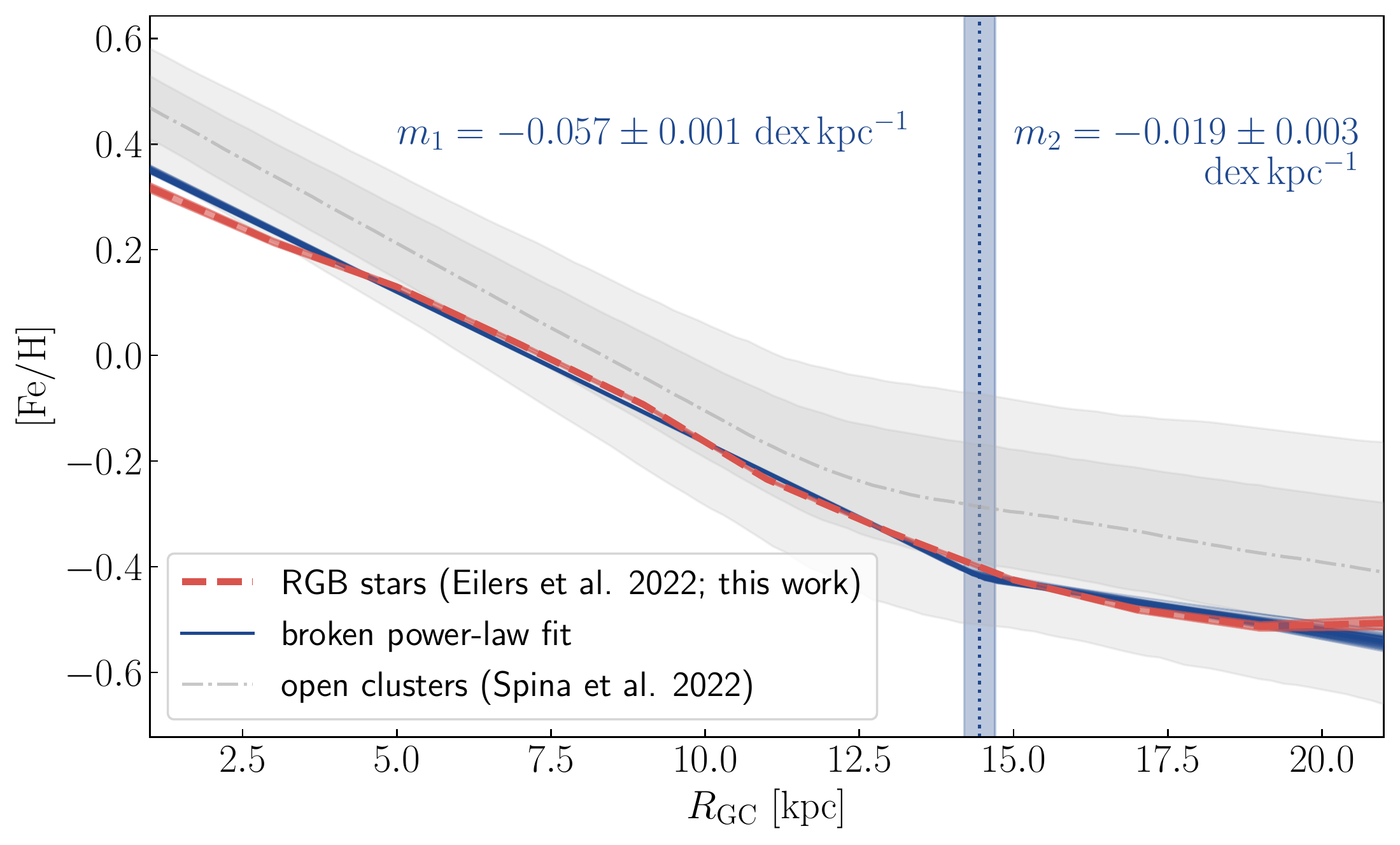}
    \caption{The observed median [Fe/H] gradient with respect to Galactocentric radius (red) is well fitted by a broken power-law (blue) with an inner slope $m_1$ and an outer slope $m_2$. Thin blue lines indicate a $100$ random draws from the posterior probability distribution of the fit parameters. The vertical dotted line indicates the break in the power-law at $14.5\pm0.3$~kpc. The metallicity gradient derived from open clusters is shown in grey showing a steeper inner slope of $-0.064\pm0.007~\rm dex\,kpc^{-1}$ \citep{Spina2022}. If the open clusters are on average younger than the RGB stars we would expect qualitatively a flattening of the gradients with age due to radial migration \citep[e.g.][]{Frankel2019, Frankel2020}. 
    \label{fig:feh_gradient}}
\end{figure}

The median [Fe/H] abundance gradient shown in detail in Fig.~\ref{fig:feh_gradient} evolves most strongly by $\Delta{\rm [Fe/H]} =0.82$~dex between the inner and the outer part of the Galactic disk. When fitted with a broken power-law \citep[e.g.][]{Donor2020, Spina2022}, i.e.
\begin{align}
    {\rm [Fe/H]} = \begin{cases}
    m_1\cdot R_{\rm GC} + b~~~~~~~~~~~~~~~~~~~~~~~& {\rm if}~R_{\rm GC}\leq k\\
    (m_1\cdot k +b) + m_2 \cdot (R_{\rm GC} - k)~~~~& {\rm if}~R_{\rm GC}> k\\
    \end{cases}
\end{align}
using the Monte Carlo Markov Chain implementation \texttt{emcee} \citep{emcee} to fit for the inner and outer slopes $m_1$ and $m_2$, the maximum metallicity at the Galactic center $b$, as well as the break (``knee'') in the power-law $k$, we find an inner slope of $m_1=-0.057\pm0.001~\rm dex\,kpc^{-1}$ and an outer slope of $m_2=-0.019\pm0.003~\rm dex\,kpc^{-1}$. The break in the power-law appears at $k=14.5\pm0.3~\rm kpc$, and the y-intercept is $b=0.41\pm0.01~\rm dex$. 

While we see strong abundance gradients in the low-$\alpha$ disk, the high-$\alpha$ disk stars are more chemically mixed and uniformly distributed across the disk, confirming that they are predominantly older than the low-$\alpha$ stars \citep[e.g.][]{BovyRix2016, Mackereth2019, GandhiNess2019, Buck2020b, Lu2021}. 

While most abundance trends are radially symmetric with respect to the Galactic center, we do find some abundances that do not show radial symmetry. For instance there seems to be an anomalously high [Co/Fe] feature in the Southern part of the disk, and parts of the outer disk seem to have a lower [Na/Fe] abundance (see Fig.~\ref{fig:maps_lo}). We do currently not know whether these are real physical effects or simply additional, non-accounted for systematic issues with the data. 

\begin{sidewaysfigure}[!t]
    \centering
    \includegraphics[width=\textwidth]{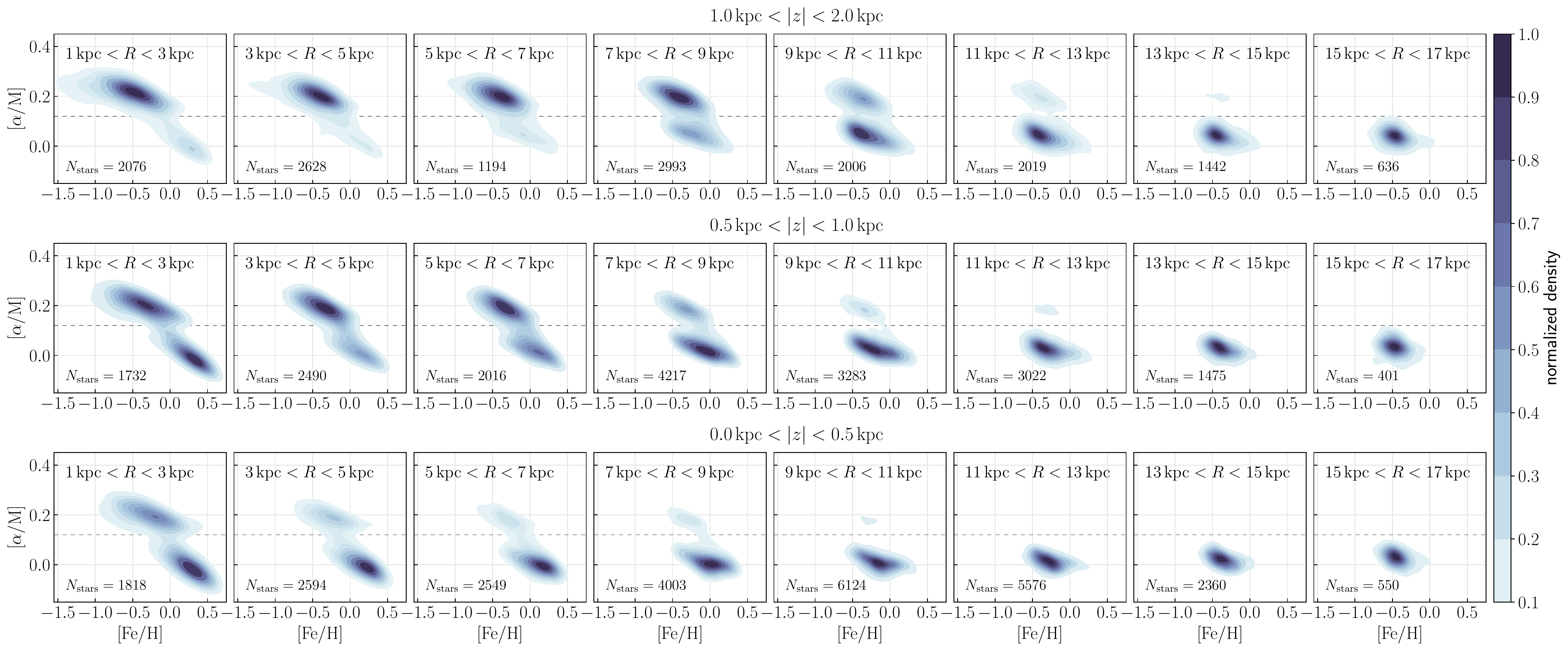}
    \caption{The observed metallicity distribution [$\alpha$/M] vs. [Fe/H] of stars within the Milky Way disk split by Galactocentric radius $R$ and height above the Galactic plane $|z|$. Note that for this plot, we included all stars within $|z|<2$~kpc. The horizontal dashed line indicates the split between low- and high-$\alpha$ stars at [$\alpha$/M]$\pm 0.12$. 
    \label{fig:H15}}
\end{sidewaysfigure}

\subsection{Metallicity Bimodality}

Making use of the newly self-calibrated abundances we can now review the metallicity distribution of stars within the Milky Way disk. The distribution of stars in the [$\alpha$/M] vs. [Fe/H] plane has long been an important  diagnostics of chemical evolution, since [$\alpha$/M] provides a rough chemical ``clock'' due to the different timescales of core collapse supernovae and supernovae Ia enrichment \citep[e.g.][]{Tinsley1979, MatteucciGreggio1986, McWilliam1997}. 

The observed bimodality of the low- and high-$\alpha$ sequence is clearly visible both within the Galactic disk as well as inside the bulge at $R<5$~kpc. In the outskirts of the disk, i.e. at $R\gtrsim 13$~kpc, most stars lie within the low-$\alpha$ sequence, which is consistent with previous results \citep[e.g.][]{Bensby2011, Nidever2014, Hayden2015, Schultheis2017, Griffith2020, Queiroz2020_starhorse}. 

\begin{sidewaysfigure}[!t]
    \centering
    \includegraphics[width=\textwidth]{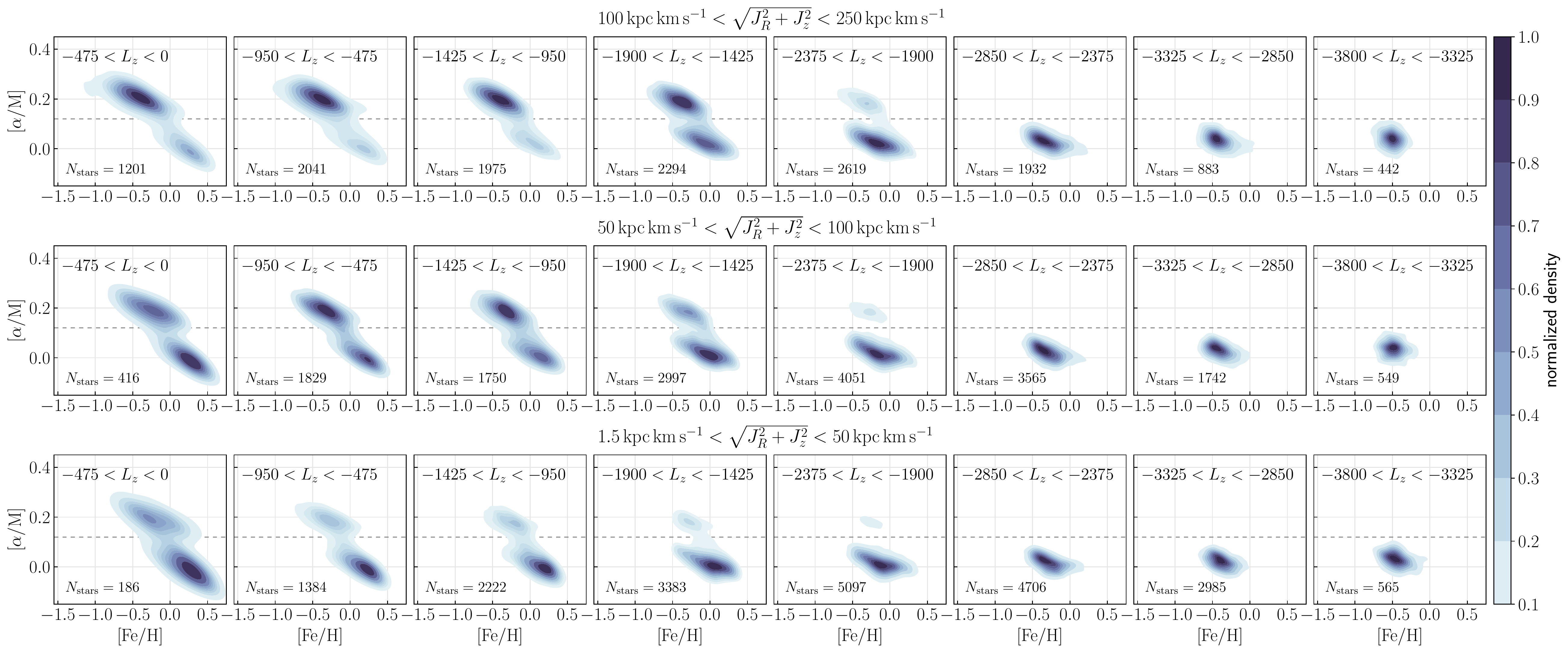}
    \caption{Analogous to Fig.~\ref{fig:H15}, but the disk stars are split by their orbital actions $L_z$, denoting their angular momentum, and $\sqrt{J_R^2+J_z^2}$, indicating the circularity and co-planarity of the stellar orbits. The observed bimodality in the stellar metallicity distribution is more pronounced when split by orbital actions instead of Galactocentric radius and height. 
    \label{fig:H15_actions}}
\end{sidewaysfigure}

This bimodality is even more prominent when dividing the stars in bins of orbital actions rather than by position. Fig.~\ref{fig:H15_actions} shows the same metallicity distribution of Milky Way disk stars split by the stars' angular momentum $L_z$, and their vertical and radial action $\sqrt{J_R^2+J_z^2}$, indicating how circular and co-planar the stellar orbits are. In most panels the two peaks of the metallicity distribution are stronger peaked and more compact compared to Fig.~\ref{fig:H15}. \\

\section{Comparison to previous studies}

\subsection{Distance estimates and abundance maps}\label{app}

\begin{figure}
    \centering
    \includegraphics[width=\textwidth]{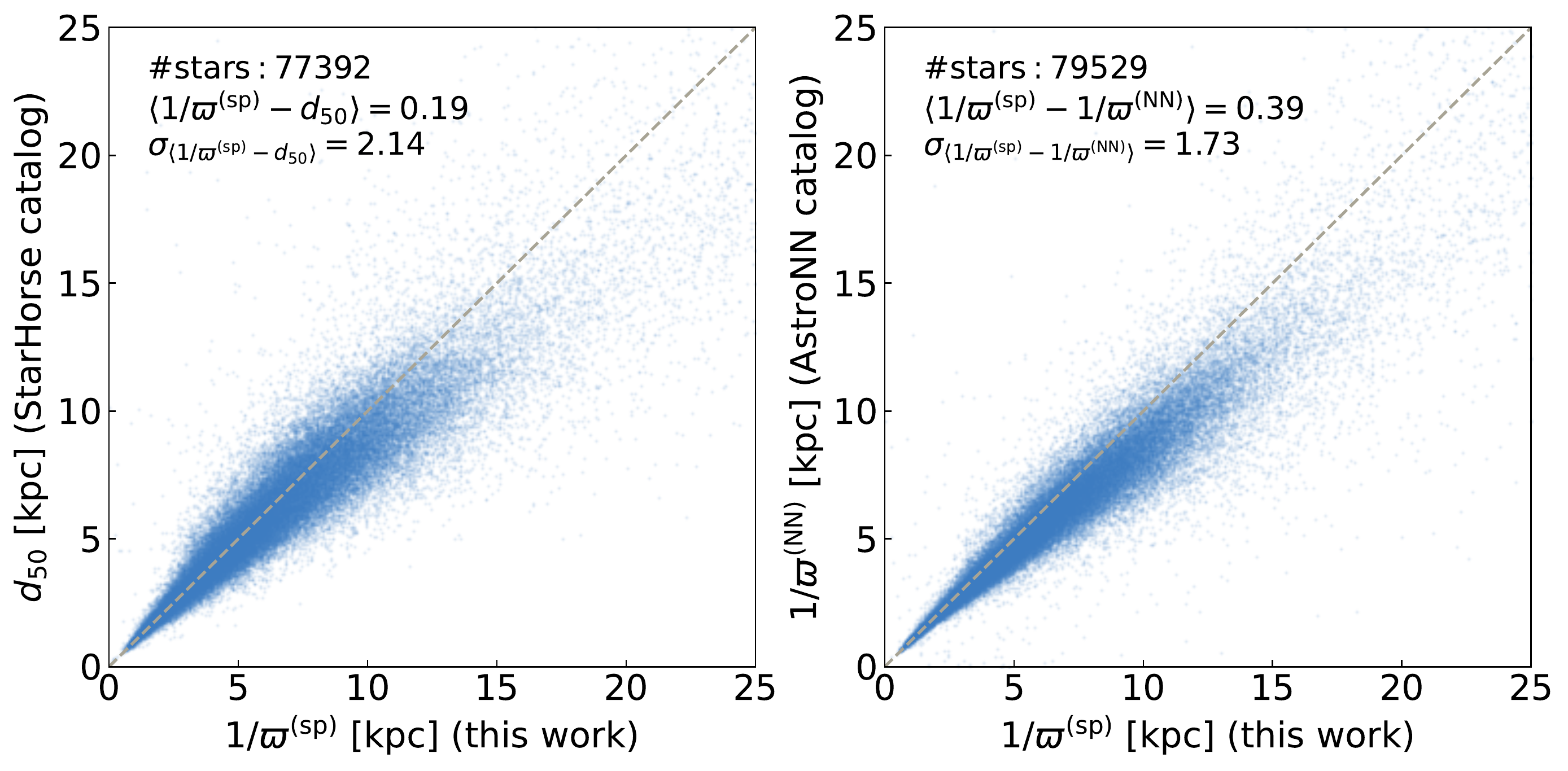}
    \caption{Comparison of our spectrophotometric distances to \texttt{StarHorse} stellar distances (\textit{left}) and \texttt{AstroNN} distance estimates (\textit{right}). Both catalogs differ from our estimated distances at small parallaxes and large distances. We discuss these biases and differences in the text. }
    \label{fig:dist2}
\end{figure}

In Fig.~\ref{fig:dist2} we compare our spectrophotometric estimates to previous work, such as the \texttt{StarHorse} catalog by \citet{Queiroz2020_starhorse}, who determined distances to stars using an isochrone-fitting approach, as well as the \texttt{AstroNN} catalog by \citet{LeungBovy2019b}, who apply a neural net to derive parallaxes as well as chemical abundances from the stellar spectro-photometric features. We find reasonable agreement with both other studies, although we generally predict larger distances compared to previous work. 
This effect could potentially be explained, if (implicit) cuts on parallaxes are performed when constructing the training sets, since the removal of negative parallax estimates biases the training set to more nearby stars thus leading to an underestimation of the distances \citep{HER2019}. 

\begin{figure}[!t]
    \centering
    \includegraphics[width=.5\textwidth]{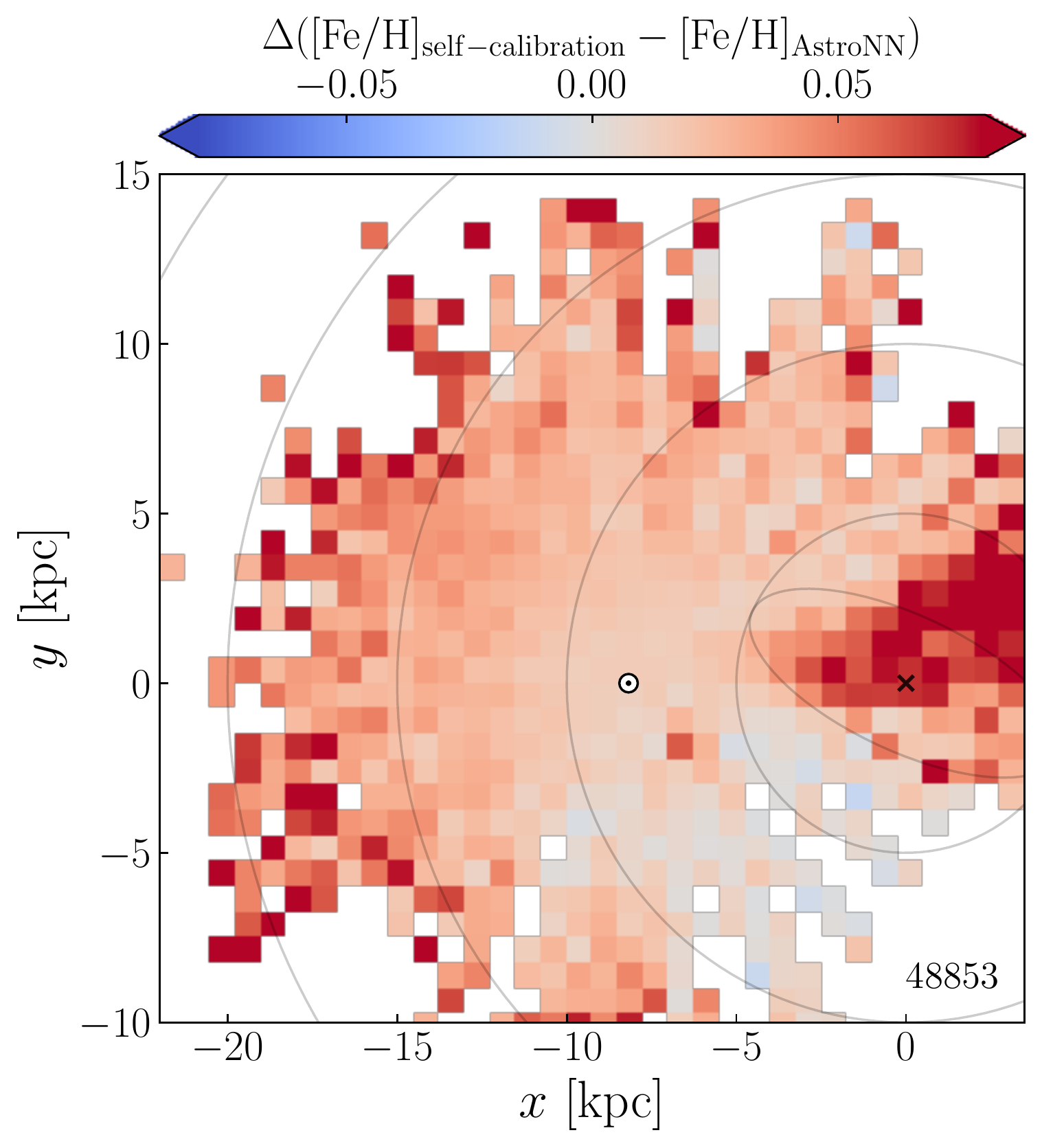}
    \caption{Map of the Galactic disk colored by the difference in the estimated metallicity [Fe/H] of our model and the AstroNN model introduced by \citet{LeungBovy2019b}. Discrepancies are visible between the two studies within the Galactic bulge region at $R<5$~kpc. Note, however, that the interpretation of this figure is difficult, since our method provides only a gradient in the abundances and no absolute calibration. The number in the lower right corner denotes the number of stars in the plot. 
    \label{fig:diff}}
\end{figure}

In Fig.~\ref{fig:diff} we compare our self-calibrated metallicity [Fe/H] to metallicities derived by the \texttt{AstroNN} algorithm \citep{LeungBovy2019a}. 
The authors had seen complex abundance patterns in the Milky Way disk, which we do not observe (see Fig.~\ref{fig:maps_lo} or \ref{fig:maps_hi}). The comparison in Fig.~\ref{fig:diff} shows the discrepancies between the results of the two approaches, which are particularly prominent towards the Galactic center.  

These trends could potentially be explained due to differences in the observed population of stars towards the Galactic bulge compared to the disk due to the \apogee\ selection function. The selection function is shallower in the Galactic center, and therefore only the most luminous red giant stars are observed in the bulge \citep[e.g.][]{Zasowski2017, apogeeDR16}. Hence, the $\log g$ distribution of the targeted stars differs between the Galactic bulge and disk, resulting in an average $\log g$ that is significantly higher towards the Galactic center. Thus it is possible that selection effects could explain the differences observed in Fig.~\ref{fig:diff}, which would also provide an explanation for the chemically distinct stellar population in the bulge reported in \citet{Bovy2019} and \citet{LeungBovy2019b}. 

\subsection{On-bar vs. off-bar stellar abundances}\label{app:lian}
\begin{figure}
    \centering
    \includegraphics[width=.6\textwidth]{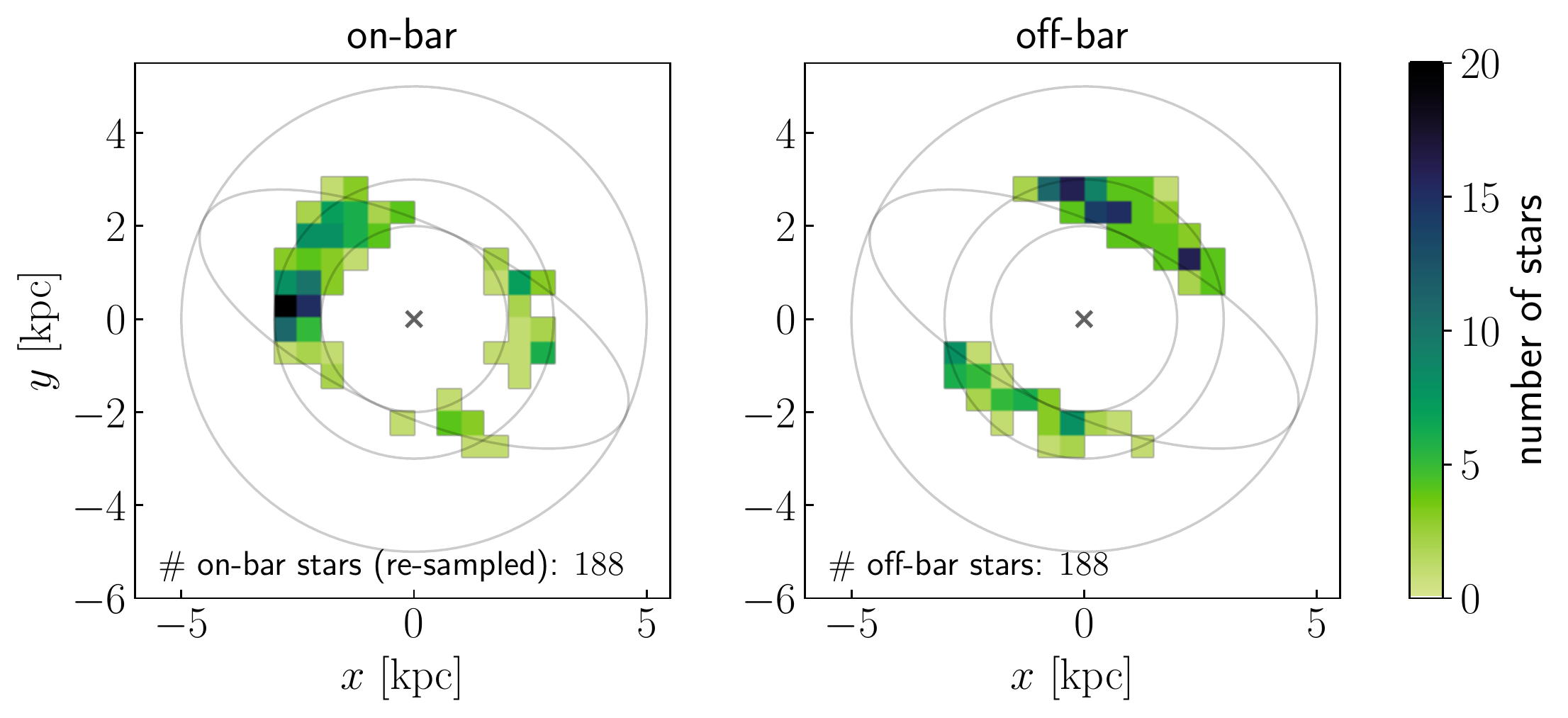}
    \includegraphics[width=.39\textwidth]{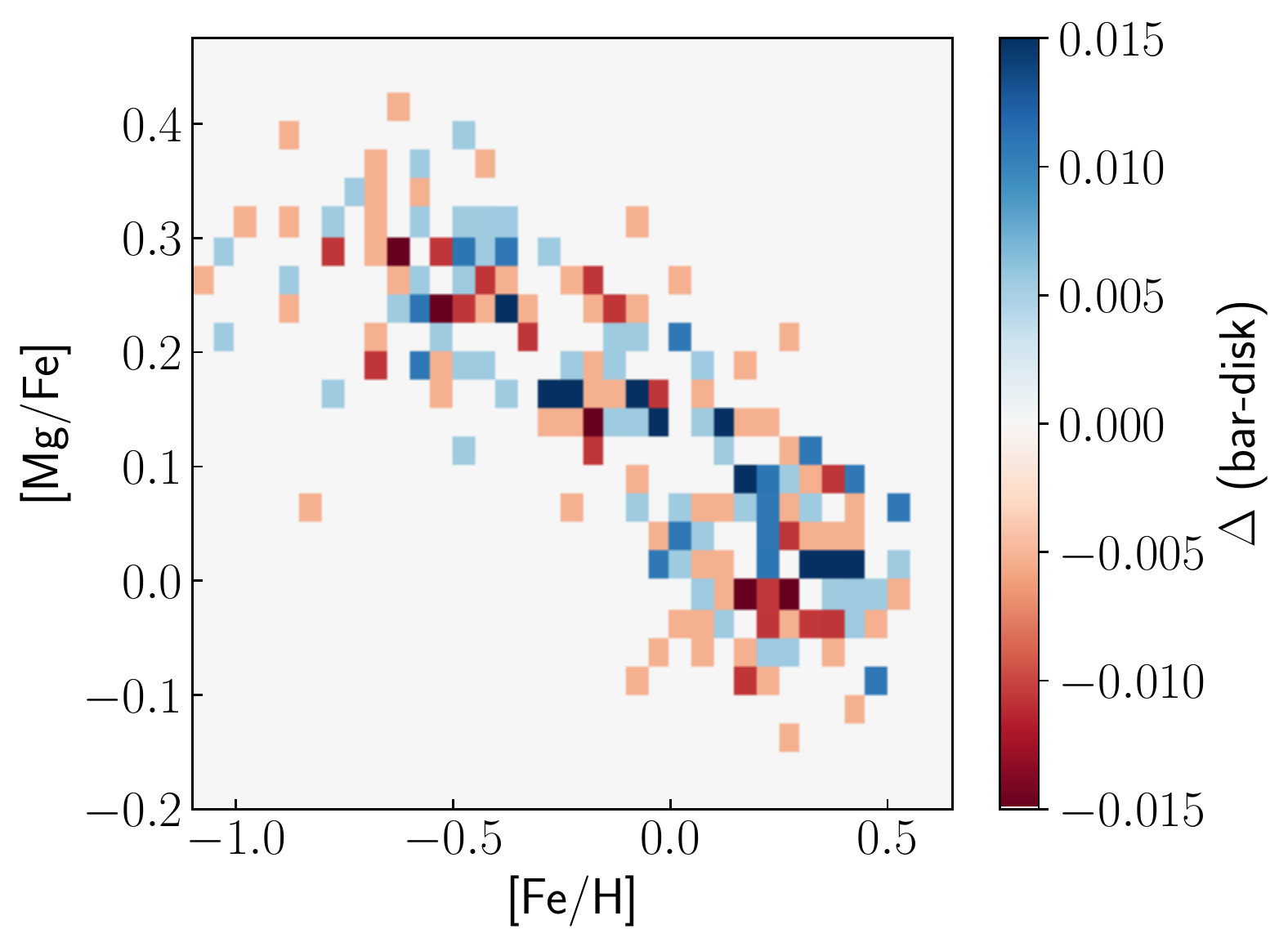}
    \caption{Comparison between on- and off-bar stars within the bulge, i.e. $|z|<0.5$~kpc. \textit{Left panels:} Spatial distribution between on- and off-bar stars. Galactocentric rings at $2, 3$ and $5$~kpc are shown as grey thin lines, as well as the outline of the Galactic bar. \textit{Right panel:} The [Mg/Fe] vs. [Fe/H] distributions of the two stellar samples do not show any significant differences. }
    \label{fig:Lian_comparison}
\end{figure}
Inspired by the results reported in \citet{Lian2021} we also study the [Mg/Fe] vs.\ [Fe/H] distribution of stars on the bar compared to stars outside the bar within the Milky Way bulge at $R<3$~kpc. Since the number and spatial distribution of on- and off-bar stars varies significantly, we follow the method described in \citet{Lian2021} and split the parameter space into bins of $\Delta R = 0.5$~kpc and $\Delta z = 0.25$~kpc. We then select a random subset of on-bar stars corresponding to the number of off-bar stars in each spatial bin. This procedure ensures the same number and approximate spatial distribution of stars in both the on- and off-bar samples. In practice this selects stars within a very narrow annulus, i.e.\ $2~{\rm kpc}<R_{\rm GC}<3~\rm kpc$ (see left panels of Fig.~\ref{fig:Lian_comparison}). 
In contrast to \citet{Lian2021}, we do not find any significant difference in the chemical properties between on- and off-bar stars, as shown in the right panel of Fig.~\ref{fig:Lian_comparison}. These results, however, seem to be independent from our self-calibration; that is, they are not produced by abundance systematics or evolution. Thus the main difference between the two analyses is different estimates in the stellar parallaxes, which might be causing the observed discrepancies. Additionally, the measurement is very noisy due to the very small number of stars (i.e. $188$) considered for this comparison. Hence, if there is indeed a chemical abundance difference between stars on the bar and off the bar, the effect is very small.

\section{Summary \& Discussion}

In this paper we construct a data-driven model to self-calibrate all $20$ stellar abundances of RGB stars within the Milky Way disk and bulge reported in \apogee\ DR16, in order to remove any effects on the nuisance parameter surface gravity $\log g$. The dependency on $\log g$ presumably is a combination of both imperfections in the data reduction, calibration, and modeling, as well as expected effects due to stellar evolution along the red giant branch. Neither effect, however, tracks changes in the stars' birth material composition. By applying our self-calibration model we aim to remove all effects and hence calibrate all abundances to the same stellar evolutionary state, just above the red clump. 

The self-calibration approach works well for most elements where the applied self-calibration completely eliminates or significantly reduces any trends with $\log g$ (e.g. Fig.~\ref{fig:feh} and \ref{fig:mgfe}). 
However, some effects remain in particular in regions with very small or Galactocentric radii where the number of stars in the training set decreases, as well as for some elements (e.g. [Al/Fe], see Fig.~\ref{fig:alfe}) our model does not remove all dependencies on $\log g$. This could be due to the limited flexibility of our model or due to other parameters that systematically bias the reported abundances. 
For [Al/Fe] in particular, \citet{Jonsson2020} reported an unexplained observed bimodality in the distribution of abundances for cool giant stars, i.e. $T_{\rm eff}<4000$~K. While an intrinsic bimodality would not bias our model, a bimodality that is caused by systematic errors due to calibration or data reduction problems, would cause artificial structure which could affect our results.

We find that all stellar abundances evolve smoothly with Galactocentric radius $R_{\rm GC}$, as well as with the height above the Galactic plane $z_{\rm GC}$. The abundances vary similarly smoothly with respect to the orbital actions $J_R$, $J_z$, and $L_z$. 
Thus, we find no evidence that stars in the bulge and bar are chemically distinct from disk stars and thus likely originate from the same stellar population. 
This implies that likely similar nucleosynthetic processes enriched both bar and disk stars despite the different star formation histories and physical conditions of these regions \citep[e.g.][]{Griffith2020, Ness2021}. 

We observe the strongest abundance gradient in [Fe/H], which varies by more than $0.8$~dex between the inner and the outer parts of the Galactic disk. When fitted with a broken power-law we find a slope of $m_1=-0.057\pm0.001~\rm dex\,kpc^{-1}$ in the inner part and a shallower slope of $m_2=-0.019\pm0.003~\rm dex\,kpc^{-1}$ beyond the break at $14.5\pm0.3~\rm kpc$. The break in the power-law that we measure occurs at larger radii than observed in the metallicity gradient of open clusters, which indicate a break around $12-14$~kpc \citep[e.g.][]{Donor2020, Spina2022}. Our measured inner slope of the metallicity gradient of $-0.057\pm0.001~\rm dex\,kpc^{-1}$ is shallower than the slope derived for open clusters, i.e. $-0.064\pm0.007~\rm dex\,kpc^{-1}$ \citep{Spina2022}, and also shallower than the metallicity gradient measured for red clump stars, i.e. $-0.073\pm0.002~\rm dex\,kpc^{-1}$ \citep{Frankel2019}. One would expect qualitatively a flattening of the gradient with age due to radial migration \citep{Frankel2019, Frankel2020}. 

The observed smooth gradients across all chemical abundances are maybe unsurprising, considering that all elemental abundances are closely linked with each other \citep[e.g.][]{Ness2019, Ness2021}, possibly even hewing closely to a two- or three-dimensional surface. However, each element comprises some additional unique information \citep[e.g.][]{Weinberg2021, TingWeinberg2021}. 
Thus it would clearly be interesting to study any discrepancies in the spatial distribution of stellar abundances that would deviate from the mean metallicity gradients. However, this is beyond the scope of this paper and could be analyzed in detail in future work.
The full table of abundances can be downloaded\footnote{\href{https://doi.org/10.5281/zenodo.5963324}{https://doi.org/10.5281/zenodo.5963324}}. 

While we see strong gradients in the stellar abundances for low-$\alpha$ stars, high-$\alpha$ stars are more chemically mixed and do not exhibit any strong gradients across the Milky Way disk. 
The observed smooth evolution of the abundances is in good agreement with \citet{Griffith2020}, who found nearly identical metallicity trends between bulge and disk stars after accounting for differences in the mean $\log g$ between the different stellar populations \citep[see also][]{Bensby2017, Weinberg2019, Weinberg2021}. 

However, our results are in strong contrast to the complex abundance pattern across the Milky Way disk reported in \citet{LeungBovy2019b} and the reported chemically distinct stellar population in the Galactic bulge and bar \citep[][]{Bovy2019, LeungBovy2019a, Queiroz2020}. Such trends could potentially appear (but be spurious) if the \apogee\ selection function has not been accounted for, since the selection function is different towards the Galactic bulge compared to the disk. Thus the $\log g$ distribution of the stars targeted by \apogee\ differs between the Galactic bulge and disk resulting in an average $\log g$ that is significantly higher towards the Galactic center \citep[e.g.][]{Zasowski2017, apogeeDR16}. Hence systematic errors in the stellar abundances that correlate with $\log g$ can cause artificial differences between the bulge and disk abundances \citep[e.g.][]{Weinberg2019, Griffith2020, Santos-Peral2020}, which will lead to biases if such selection effects are not accounted for.
Our self-calibration method applied here implicitly accounts for such selection effects by effectively correcting all stars to the same evolutionary state, i.e. the same value of $\log g$. 

Our results also seem to be in tension with the reported chemically distinct properties in the Galactic bar and disk found by \citet{Lian2021}. When following the method described in \citet{Lian2021} to select on- and off-bar stars we do not find any significant differences between the bar and the surrounding disk in the [Mg/Fe] vs.\ [Fe/H] abundances (Fig.~\ref{fig:Lian_comparison}). We speculate that differences in the stellar distance estimates could be causing the observed effect. This supports the evidence that the stellar populations within the bar are not significantly chemically distinct from the Milky Way disk.

The pursuit of linking stars to their birth places using their current day abundances \citep{FreemanBland-Hawthorn2002, Ness2021, Bridget2021} in practice asks about \textit{birth abundances}. 
Because our sample included only the top of the red-giant branch (above the red clump), we calibrated the abundances to what might be thought of as a ``reference'' evolutionary state at $\log g=2.2$. Since the abundances of interest to chemodynamics or chemical tagging are really the birth abundances, it would be interesting and valuable to calibrate the measured abundances back to birth abundances or pre-main-sequence abundances. This is not possible with the chosen sample of RGB stars, but would be possible with a more extended data set including also main sequence stars. Such a calibration could open up new opportunities for measuring and understanding the formation and evolution of the Milky Way. 
We also expect that new capabilities would arise if we had precise age measurements for the stars in our sample, since at least some element abundances are expected to vary monotonically with stellar age \citep{Martig2016}; however at present, age estimates for these stars are not sufficiently precise \citep{Ness2016b}. 

A generalization of this self-calibration approach could also enable us in the future to cross calibrate among different surveys that have different selection functions and hence different biases---and stars in different evolutionary states---in their reported stellar abundances.

\acknowledgements
It is our great pleasure to thank David Weinberg, Michael Blanton, Eric Bell and the astronomical data group at the Flatiron Institute for helpful discussions. 

ACE acknowledges support by NASA through the NASA Hubble Fellowship grant $\#$HF2-51434 awarded by the Space Telescope Science Institute, which is operated by the Association of Universities for Research in Astronomy, Inc., for NASA, under contract NAS5-26555. 

Funding for the Sloan Digital Sky 
Survey IV has been provided by the 
Alfred P. Sloan Foundation, the U.S. 
Department of Energy Office of 
Science, and the Participating 
Institutions. 

SDSS-IV acknowledges support and 
resources from the Center for High 
Performance Computing  at the 
University of Utah. The SDSS 
website is www.sdss.org.

SDSS-IV is managed by the 
Astrophysical Research Consortium 
for the Participating Institutions 
of the SDSS Collaboration including 
the Brazilian Participation Group, 
the Carnegie Institution for Science, 
Carnegie Mellon University, Center for 
Astrophysics | Harvard \& 
Smithsonian, the Chilean Participation 
Group, the French Participation Group, 
Instituto de Astrof\'isica de 
Canarias, The Johns Hopkins 
University, Kavli Institute for the 
Physics and Mathematics of the 
Universe (IPMU) / University of 
Tokyo, the Korean Participation Group, 
Lawrence Berkeley National Laboratory, 
Leibniz Institut f\"ur Astrophysik 
Potsdam (AIP),  Max-Planck-Institut 
f\"ur Astronomie (MPIA Heidelberg), 
Max-Planck-Institut f\"ur 
Astrophysik (MPA Garching), 
Max-Planck-Institut f\"ur 
Extraterrestrische Physik (MPE), 
National Astronomical Observatories of 
China, New Mexico State University, 
New York University, University of 
Notre Dame, Observat\'ario 
Nacional / MCTI, The Ohio State 
University, Pennsylvania State 
University, Shanghai 
Astronomical Observatory, United 
Kingdom Participation Group, 
Universidad Nacional Aut\'onoma 
de M\'exico, University of Arizona, 
University of Colorado Boulder, 
University of Oxford, University of 
Portsmouth, University of Utah, 
University of Virginia, University 
of Washington, University of 
Wisconsin, Vanderbilt University, 
and Yale University. 

\software{numpy \citep{numpy}, scipy \citep{scipy}, matplotlib \citep{matplotlib}, astropy \citep{astropy}, gala \citep{adrian_price_whelan_2020_4159870}}

\appendix

\section{Abundance Maps for all elements}\label{app:all_maps}

\begin{sidewaysfigure}
    \centering
    \includegraphics[width=\textwidth]{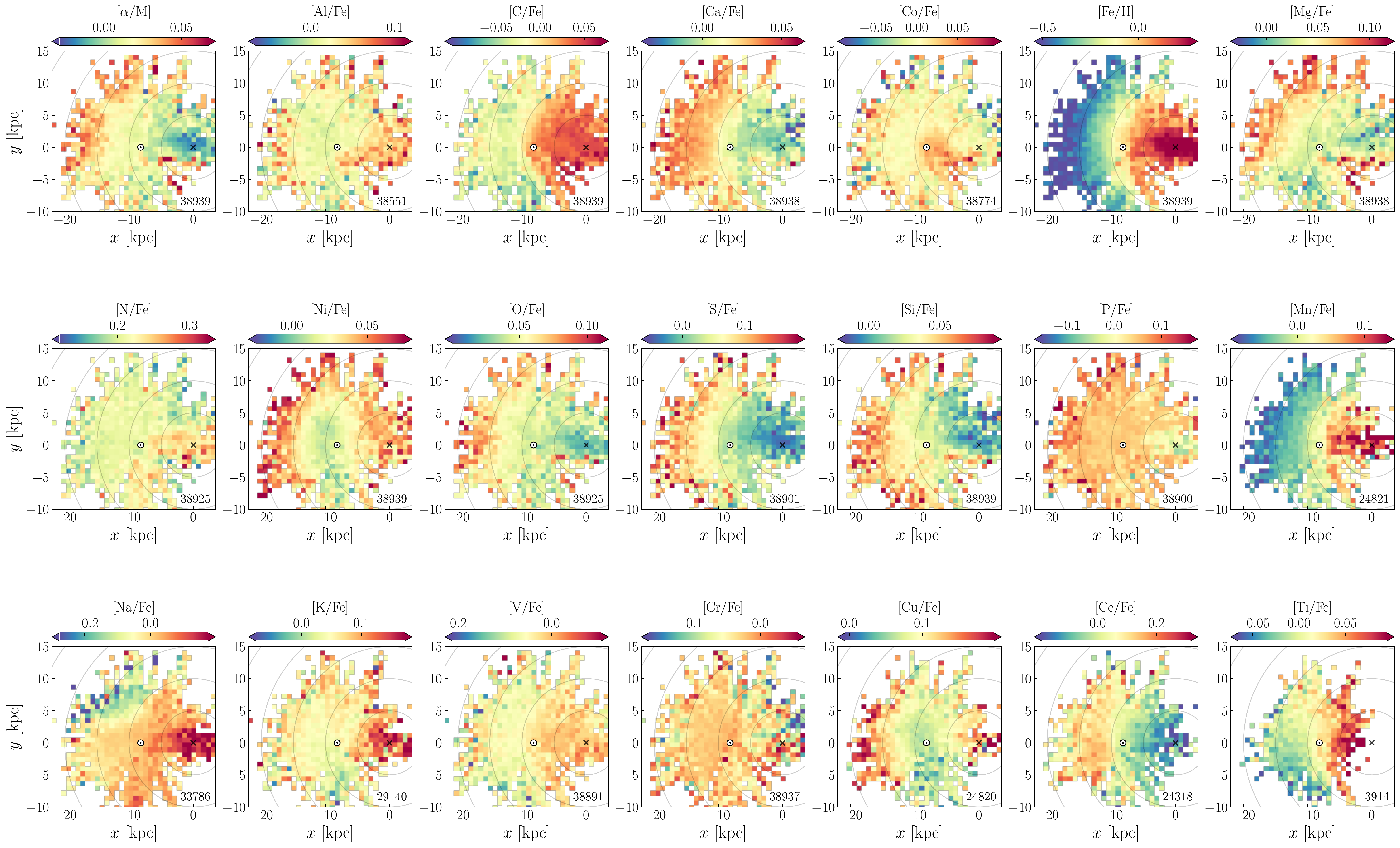}
    \caption{Maps of the Galactic disk (within $|z|<1~\rm kpc$ or a $6^\circ$ wedge) colored by different chemical abundances after applying the self-calibration showing only low-$\alpha$ stars, i.e. $[\alpha/{\rm M}] < 0.12$. We show the median abundances of all stars in spatial bins of $\Delta x = \Delta y = 0.75$~kpc, whenever a bin contains more than $3$ stars. The colorbar ranges between the $7$th to $93$rd percentile of median abundances in each panel. 
    \label{fig:maps_lo}}
\end{sidewaysfigure}

\begin{sidewaysfigure}
    \centering
    \includegraphics[width=\textwidth]{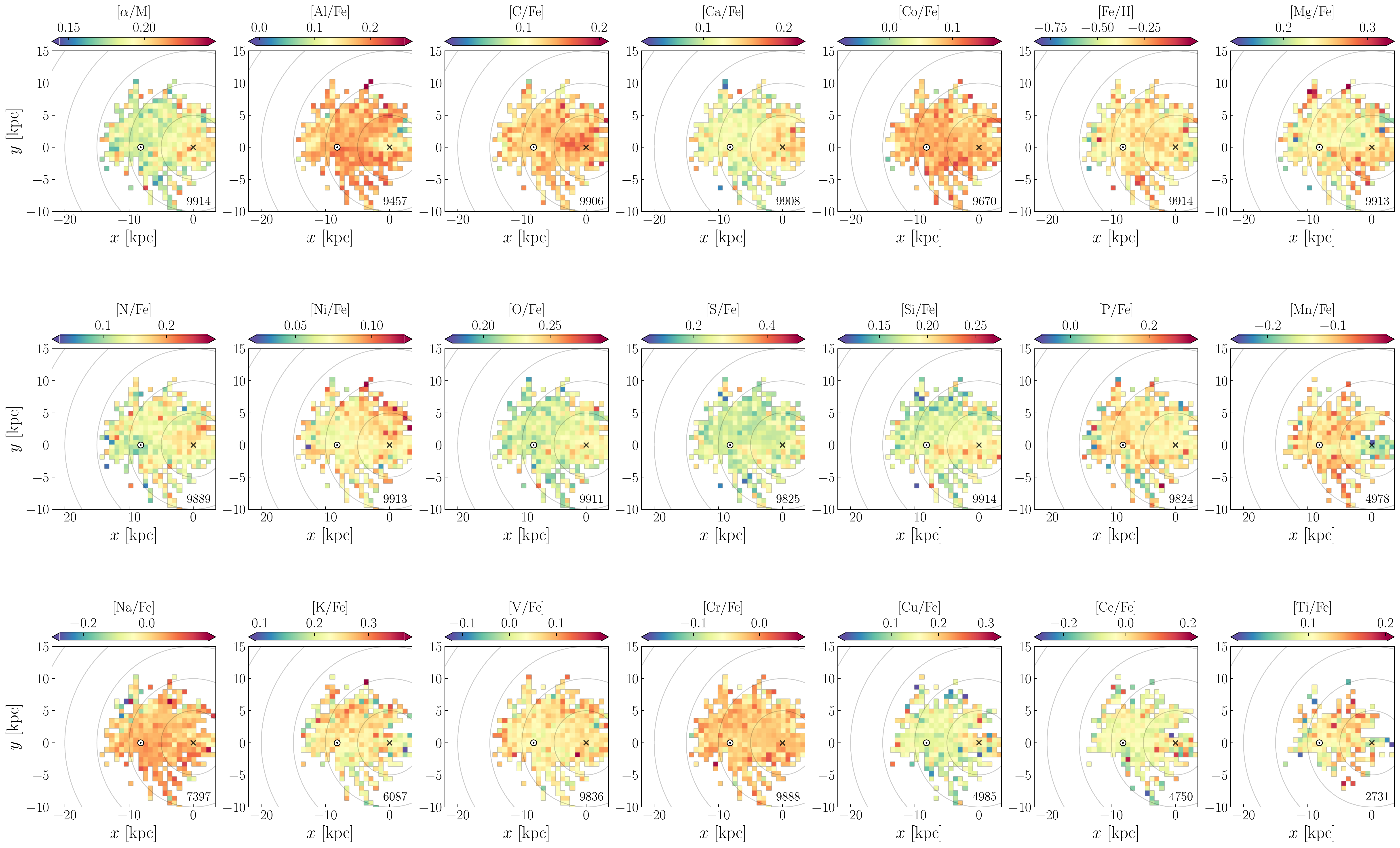}
    \caption{Same as Fig.~\ref{fig:maps_lo} for high-$\alpha$ stars, i.e. $[\alpha/{\rm M}] > 0.12$. 
    \label{fig:maps_hi}}
\end{sidewaysfigure}

Galactic maps of all self-calibrated abundances of RGB stars for abundances reported by \apogee\ DR16 are shown in Fig.~\ref{fig:maps_lo} and \ref{fig:maps_hi} split between low-$\alpha$ and high-$\alpha$ stars, respectively. We also report all median abundance gradients with respect to Galactocentric radius in Tab.~\ref{tab:gradients} to \ref{tab:gradients3}. 

\begin{deluxetable*}{CRRRRRRR}
\setlength{\tabcolsep}{3pt}
\tablecaption{Radial gradients of stellar abundances. \label{tab:gradients}}
\tablehead{\dcolhead{R_{\rm GC}} & \dcolhead{{\rm[\alpha/M]}+\Delta}& \dcolhead{{\rm [C/Fe]}+\Delta}& \dcolhead{{\rm [N/Fe]}+\Delta}& \dcolhead{{\rm [O/Fe]}+\Delta}& \dcolhead{{\rm [Na/Fe]}+\Delta}& \dcolhead{{\rm [Mg/Fe]}+\Delta}& \dcolhead{{\rm [Al/Fe]}+\Delta}\\
\colhead{[kpc]} & & & & }
\startdata
1  &  -0.015\pm0.002  &  0.047\pm0.002  &  0.240\pm0.004  &  0.023\pm0.002  &  0.156\pm0.010  &  0.019\pm0.002  &  0.048\pm0.004  \\ 
3  &  -0.003\pm0.001  &  0.041\pm0.001  &  0.235\pm0.003  &  0.032\pm0.001  &  0.090\pm0.006  &  0.022\pm0.002  &  0.044\pm0.003  \\ 
5  &  0.004\pm0.001  &  0.039\pm0.001  &  0.225\pm0.002  &  0.038\pm0.001  &  0.054\pm0.004  &  0.030\pm0.001  &  0.027\pm0.002  \\ 
7  &  0.011\pm0.001  &  0.027\pm0.001  &  0.214\pm0.002  &  0.043\pm0.001  &  0.029\pm0.003  &  0.035\pm0.001  &  0.021\pm0.001  \\ 
9  &  0.013\pm0.001  &  0.008\pm0.001  &  0.209\pm0.001  &  0.044\pm0.001  &  0.011\pm0.002  &  0.032\pm0.001  &  0.012\pm0.001  \\ 
11  &  0.016\pm0.001  &  -0.022\pm0.001  &  0.209\pm0.001  &  0.048\pm0.001  &  -0.008\pm0.002  &  0.035\pm0.001  &  0.011\pm0.001  \\ 
13  &  0.021\pm0.001  &  -0.032\pm0.001  &  0.208\pm0.001  &  0.054\pm0.001  &  -0.019\pm0.003  &  0.046\pm0.001  &  0.014\pm0.001  \\ 
15  &  0.029\pm0.001  &  -0.025\pm0.002  &  0.207\pm0.002  &  0.064\pm0.001  &  -0.030\pm0.005  &  0.057\pm0.001  &  0.024\pm0.002  \\ 
17  &  0.034\pm0.001  &  -0.026\pm0.003  &  0.207\pm0.003  &  0.068\pm0.002  &  -0.032\pm0.011  &  0.067\pm0.002  &  0.030\pm0.004  \\ 
19  &  0.035\pm0.002  &  -0.028\pm0.005  &  0.212\pm0.005  &  0.067\pm0.003  &  -0.031\pm0.014  &  0.070\pm0.003  &  0.024\pm0.005  \\ 
21  &  0.031\pm0.003  &  -0.027\pm0.006  &  0.217\pm0.007  &  0.061\pm0.004  &  -0.011\pm0.016  &  0.064\pm0.003  &  0.017\pm0.007  \\ 
 \enddata
\tablecomments{The columns denote the Galactocentric radius and the median self-calibrated abundances with uncertainties on the median, i.e. $1.253\,\sigma/\sqrt{N}$. Note, since we calibrate the abundances to a randomly chosen stellar evolutionary state of $\log g=2.2$, we provide no absolute abundances but rather gradients, and hence all values could be shifted by a small offset $\Delta$. }
\end{deluxetable*}

\begin{deluxetable*}{CRRRRRRR}
\setlength{\tabcolsep}{3pt}
\tablecaption{Radial gradients of stellar abundances. \label{tab:gradients2}}
\tablehead{\dcolhead{R_{\rm GC}} & \dcolhead{{\rm[Si/Fe]}+\Delta}& \dcolhead{{\rm [P/Fe]}+\Delta}& \dcolhead{{\rm [S/Fe]}+\Delta}& \dcolhead{{\rm [K/Fe]}+\Delta}& \dcolhead{{\rm [Ca/Fe]}+\Delta}& \dcolhead{{\rm [Ti/Fe]}+\Delta}& \dcolhead{{\rm [V/Fe]}+\Delta}\\
\colhead{[kpc]} & & & & }
\startdata
1  &  0.001\pm0.002  &  -0.011\pm0.004  &  -0.035\pm0.004  &  0.134\pm0.012  &  -0.010\pm0.002  &  0.135\pm0.026  &  -0.007\pm0.004  \\ 
3  &  0.008\pm0.001  &  0.018\pm0.003  &  -0.015\pm0.003  &  0.116\pm0.006  &  -0.006\pm0.001  &  0.073\pm0.008  &  -0.008\pm0.003  \\ 
5  &  0.015\pm0.001  &  0.038\pm0.003  &  -0.002\pm0.002  &  0.078\pm0.003  &  -0.005\pm0.001  &  0.068\pm0.003  &  -0.007\pm0.002  \\ 
7  &  0.022\pm0.001  &  0.047\pm0.002  &  0.015\pm0.001  &  0.049\pm0.002  &  0.001\pm0.001  &  0.035\pm0.002  &  -0.018\pm0.002  \\ 
9  &  0.026\pm0.001  &  0.047\pm0.001  &  0.030\pm0.001  &  0.048\pm0.002  &  0.010\pm0.001  &  0.017\pm0.001  &  -0.036\pm0.002  \\ 
11  &  0.034\pm0.001  &  0.047\pm0.002  &  0.057\pm0.001  &  0.043\pm0.001  &  0.022\pm0.001  &  -0.002\pm0.001  &  -0.054\pm0.002  \\ 
13  &  0.040\pm0.001  &  0.049\pm0.002  &  0.074\pm0.001  &  0.047\pm0.002  &  0.029\pm0.001  &  -0.010\pm0.001  &  -0.057\pm0.002  \\ 
15  &  0.050\pm0.001  &  0.061\pm0.005  &  0.091\pm0.003  &  0.055\pm0.003  &  0.031\pm0.001  &  -0.012\pm0.002  &  -0.053\pm0.004  \\ 
17  &  0.055\pm0.002  &  0.068\pm0.009  &  0.102\pm0.005  &  0.058\pm0.005  &  0.034\pm0.001  &  -0.007\pm0.005  &  -0.050\pm0.008  \\ 
19  &  0.055\pm0.003  &  0.059\pm0.014  &  0.092\pm0.008  &  0.048\pm0.008  &  0.030\pm0.002  &  -0.019\pm0.008  &  -0.039\pm0.013  \\ 
21  &  0.049\pm0.003  &  0.058\pm0.019  &  0.103\pm0.012  &  0.041\pm0.012  &  0.028\pm0.004  &  -0.006\pm0.017  &  -0.050\pm0.015  \\ 
 \enddata
\tablecomments{Continuation of Table~\ref{tab:gradients}. }
\end{deluxetable*}

\begin{deluxetable*}{CRRRRRRR}
\setlength{\tabcolsep}{3pt}
\tablecaption{Radial gradients of stellar abundances. \label{tab:gradients3}}
\tablehead{\dcolhead{R_{\rm GC}} & \dcolhead{{\rm[Cr/Fe]}+\Delta}& \dcolhead{{\rm [Mn/Fe]}+\Delta}& \dcolhead{{\rm [Fe/H]}+\Delta}& \dcolhead{{\rm [Co/Fe]}+\Delta}& \dcolhead{{\rm [Ni/Fe]}+\Delta}& \dcolhead{{\rm [Cu/Fe]}+\Delta}& \dcolhead{{\rm [Ce/Fe]}+\Delta}\\
\colhead{[kpc]} & & & & }
\startdata
1  &  -0.014\pm0.004  &  0.134\pm0.013  &  0.317\pm0.009  &  0.000\pm0.004  &  0.045\pm0.002  &  0.129\pm0.012  &  -0.132\pm0.022  \\ 
3  &  -0.053\pm0.003  &  0.081\pm0.006  &  0.215\pm0.006  &  0.005\pm0.003  &  0.038\pm0.001  &  0.109\pm0.005  &  -0.113\pm0.012  \\ 
5  &  -0.043\pm0.002  &  0.071\pm0.003  &  0.130\pm0.005  &  0.021\pm0.002  &  0.035\pm0.001  &  0.091\pm0.003  &  -0.072\pm0.005  \\ 
7  &  -0.025\pm0.001  &  0.039\pm0.002  &  0.021\pm0.004  &  0.022\pm0.001  &  0.023\pm0.001  &  0.073\pm0.002  &  -0.027\pm0.004  \\ 
9  &  -0.021\pm0.001  &  0.006\pm0.001  &  -0.093\pm0.003  &  0.013\pm0.001  &  0.016\pm0.001  &  0.066\pm0.001  &  0.022\pm0.003  \\ 
11  &  -0.022\pm0.001  &  -0.024\pm0.001  &  -0.234\pm0.002  &  0.003\pm0.001  &  0.014\pm0.001  &  0.075\pm0.001  &  0.094\pm0.003  \\ 
13  &  -0.024\pm0.001  &  -0.040\pm0.001  &  -0.335\pm0.002  &  0.002\pm0.001  &  0.023\pm0.001  &  0.093\pm0.002  &  0.114\pm0.004  \\ 
15  &  -0.027\pm0.002  &  -0.051\pm0.002  &  -0.425\pm0.004  &  0.010\pm0.003  &  0.037\pm0.001  &  0.115\pm0.003  &  0.093\pm0.007  \\ 
17  &  -0.034\pm0.005  &  -0.068\pm0.003  &  -0.481\pm0.006  &  0.014\pm0.006  &  0.050\pm0.002  &  0.145\pm0.008  &  0.062\pm0.016  \\ 
19  &  -0.030\pm0.007  &  -0.069\pm0.004  &  -0.513\pm0.008  &  0.020\pm0.011  &  0.056\pm0.002  &  0.156\pm0.012  &  0.053\pm0.021  \\ 
21  &  -0.026\pm0.009  &  -0.065\pm0.010  &  -0.507\pm0.012  &  0.022\pm0.017  &  0.063\pm0.003  &  0.158\pm0.013  &  0.022\pm0.033  \\ 
 \enddata
\tablecomments{Continuation of Table~\ref{tab:gradients}. }
\end{deluxetable*}

\section{Comparison of abundance maps to Apogee}\label{app:apogee}

\begin{sidewaysfigure}
    \centering
    \includegraphics[width=\textwidth]{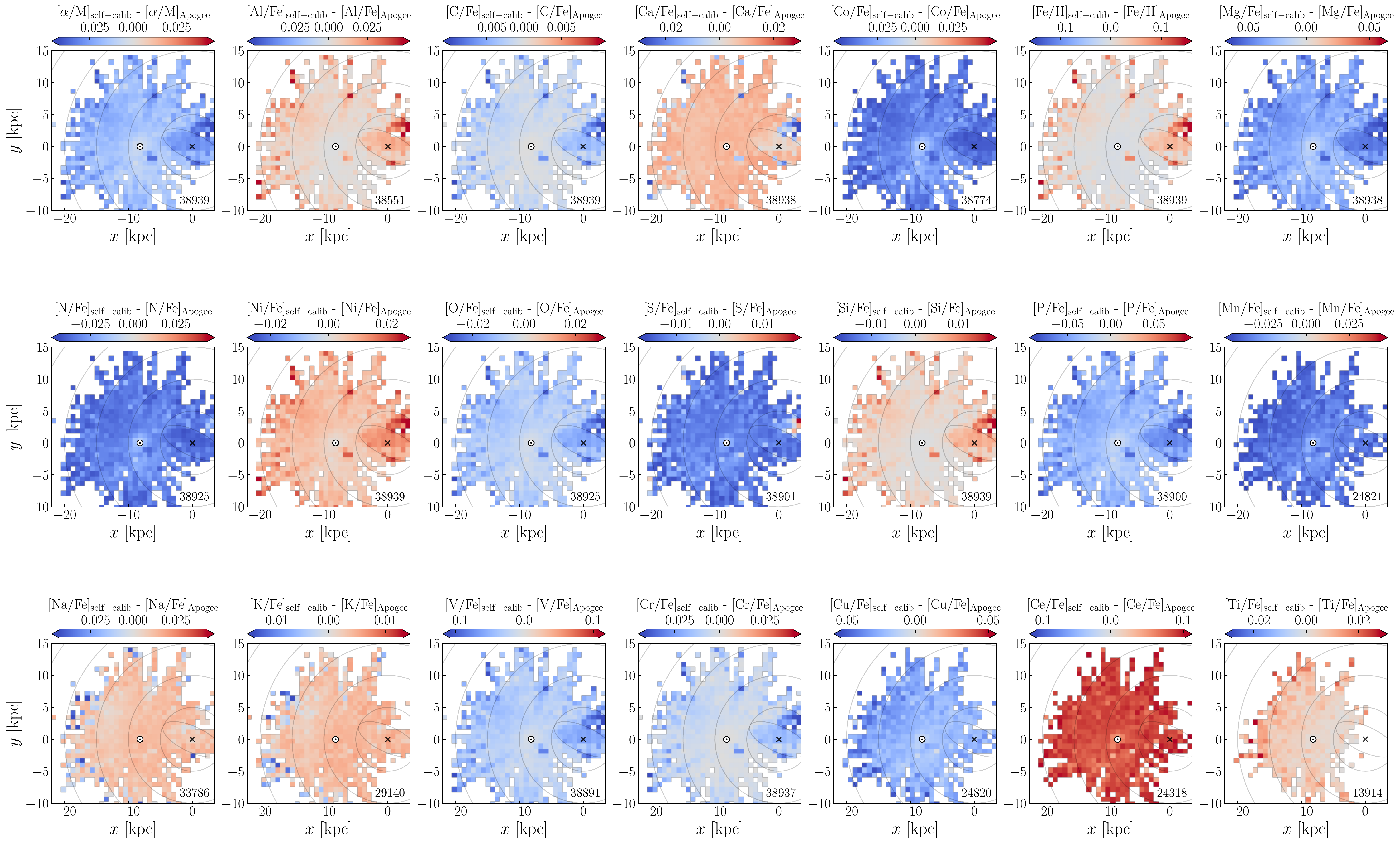}
    \caption{Galactic maps showing the differences between our self-calibrated and \apogee\ DR16 stellar abundances for low-$\alpha$ disk stars. Note, however, that the interpretation of this figure is difficult, since our self-calibration method provides only a gradient in the abundances and no absolute calibration. 
    \label{fig:maps_lo_diff}}
\end{sidewaysfigure}

\begin{sidewaysfigure}
    \centering
    \includegraphics[width=\textwidth]{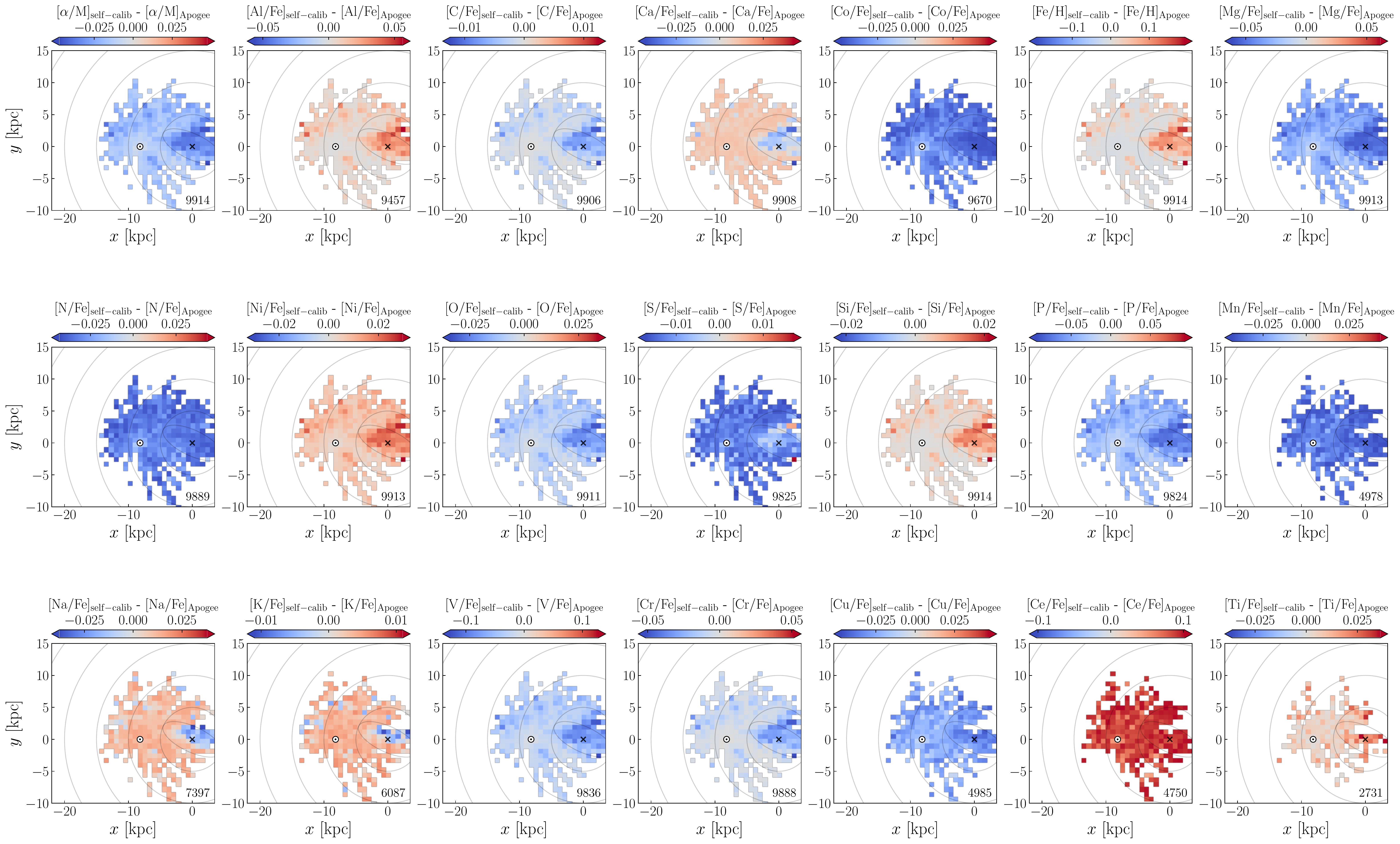}
    \caption{Same as Fig.~\ref{fig:maps_lo_diff} for high-$\alpha$ stars. 
    \label{fig:maps_hi_diff}}
\end{sidewaysfigure}

Fig.~\ref{fig:maps_lo_diff} and \ref{fig:maps_hi_diff} show Galactic maps colored by the difference between the self-calibrated stellar abundances and the abundances reported by \apogee\ DR16 for low- and high-$\alpha$ stars, respectively. The strongest discrepancies can be observed consistently for nearly all elements towards the Galactic bulge. 

\bibliography{literature_mw}

\end{document}